\documentclass[journal]{IEEEtran}

\usepackage[T1]{fontenc}
\usepackage[utf8]{inputenc}

\usepackage{amsmath}
\usepackage{amssymb}
\usepackage{amsthm}
\usepackage{mathtools}

\newtheorem{theorem}{Theorem}

\newtheorem{assumption}{Assumption}
\newtheorem{definition}{Definition}

\usepackage{algorithm}
\usepackage{algorithmicx}
\usepackage{algpseudocode}

\usepackage{graphicx}
\usepackage{subfig}
\usepackage{booktabs}
\usepackage{multirow}
\usepackage{tabularx}

\usepackage{hyperref}
\hypersetup{
    colorlinks=true,
    linkcolor=black,
    citecolor=black,
    urlcolor=blue
}

\usepackage{xcolor}

\usepackage{cite}
\usepackage{url}
\usepackage{bm}
\usepackage{bbm}

\usepackage{placeins}

\begin{document}

\title{PRIME: Plasticity Recovery in Multi-Agent Environments for UAV-Assisted Emergency Communication Networks}

\author{
  Wen~Qiu,~\IEEEmembership{Member,~IEEE,}
  Zhiqiang~He,~\IEEEmembership{Member,~IEEE,}
  Wei~Zhao,~\IEEEmembership{Member,~IEEE,}
  and~Hiroshi~Masui
  \thanks{W.~Qiu and H.~Masui are with the Department of Information and
  Communication Engineering, Kitami Institute of Technology, Japan
  (e-mail: w-qiu@ieee.org; hgmasui@mail.kitami-it.ac.jp).}
  \thanks{Z.~He is with the Graduate School of Informatics and Engineering,
  University of Electro-Communications, Japan
  (e-mail: hezhiqiang@ieee.org).}
  \thanks{W.~Zhao is with the School of Computer Science and Technology,
  Anhui University of Technology, China
  (e-mail: zhaowei@ahut.edu.cn).}
  \thanks{Corresponding authors: Wei~Zhao and Hiroshi~Masui.}
}

\markboth{IEEE Transactions on Mobile Computing}%
{Qiu \MakeLowercase{\textit{et al.}}: PRIME: Plasticity Recovery in Multi-Agent Environments}

\maketitle

\begin{abstract}
When disasters destroy terrestrial infrastructure, fleets of unmanned aerial vehicles (UAVs) can restore connectivity as temporary aerial base stations, coordinating trajectories, spectrum, and power while the mission itself keeps changing.
Most reinforcement learning controllers for these networks assume stationary conditions, and the few that handle change react to the external environment while leaving the network's internal state unexamined.
We show that sustained non-stationarity damages this internal state directly: as objectives shift, neurons progressively fall dormant and the shared policy loses the capacity to learn.
The obvious remedy, resetting dormant neurons, is unsafe under shared-parameter multi-agent training: many neurons that appear inactive are still receiving strong training gradients, and whether a neuron appears dormant depends on which agent's observations it processes.
PRIME (Plasticity Recovery In Multi-agent Environments) therefore verifies both directions before intervening.
Extending the bidirectional Silent Neuron framework to cooperative multi-agent reinforcement learning, it aggregates activation and gradient statistics over the full team batch, reads the backward signal from the gradient the training loss has already deposited , not from a hand-crafted proxy, and reinitializes only neurons that are simultaneously activation-dormant and gradient-silent.
Useful representations are preserved while learning capacity is restored.
On a phase-switching UAV emergency communication simulator, PRIME improves interquartile mean return by 24.9\% over MAPPO and holds dormant neuron fractions at 10--20\% versus 40--45\%; ablations attribute the gains to the gradient signal and team-level aggregation rather than to the specific reset operator.
A dynamic regret bound shows that the perturbation cost scales with the small silent-subspace dimension rather than the full parameter count.
\end{abstract}

\begin{IEEEkeywords}
Multi-Agent Reinforcement Learning,
Neural Plasticity,
Dormant Neuron,
UAV-Assisted Emergency Communication,
Non-Stationary Environments
\end{IEEEkeywords}

\section{Introduction}
\label{sec:introduction}

\IEEEPARstart{W}{hen} large-scale disasters strike, terrestrial communication infrastructure is often among the first casualties, and restoring connectivity in the affected area is a prerequisite for coordinating subsequent rescue and relief operations~\cite{8641424, 8660516}. Deploying multiple unmanned aerial vehicles (UAVs) as temporary aerial base stations has emerged as a practical solution, forming what is commonly referred to as a UAV-assisted emergency communication network (UAV-ECN)~\cite{8579209, 8470897}. What makes this problem distinct from other UAV deployments is that ground-level demand is driven by the evolving disaster situation rather than by a pre-designed traffic model: as rescue operations progress through distinct phases, the number, location, and priority of users shift in ways no prior distribution adequately captures. Non-stationarity is therefore not an occasional disturbance but a structural feature of the task.
 
Reinforcement learning (RL) and its multi-agent extension (MARL) have been widely adopted for UAV-ECN control, since they learn policies directly from interaction with the environment and jointly optimize trajectories, spectrum, and transmit power without requiring closed-form channel or mobility models~\cite{wu2018joint, 10.1109/JSAC.2018.2864373, 9354996, 10.1145/3555661.3560866, xing2007dynamic, xu2024trajectory}. Most of these studies, however, assume the environment is stationary or that its variation can be perceived externally. A smaller line of work addresses non-stationarity directly, through prioritized experience replay~\cite{schaul2016prioritized}, meta-RL for rapid adaptation~\cite{10.5555/3305381.3305498}, or communication-assisted phase handling such as MACPH~\cite{wang2025macph}. These methods help the agent respond to external change, but they leave open a more basic question: what happens \emph{inside} the policy network as it trains under sustained non-stationarity?

The answer matters in practice: internal degradation does not surface in standard reward curves or training loss plots, so an agent may continue to produce actions while its representational capacity quietly erodes, leaving it unable to adapt when the disaster situation transitions to a new phase. Recent work in continual reinforcement learning has documented exactly such a degradation, termed plasticity loss: as training progresses under sustained distributional shift, a growing fraction of neurons drift toward near-zero activation and become dormant, and the network's capacity to fit new data distributions steadily declines~\cite{dohare2024loss, lyle2023understanding, abbas2023loss}. The operating condition of UAV emergency communication matches this trigger exactly: demand undergoes abrupt transitions between disaster phases, and features learned for one phase become ineffective in the next.

The cooperative multi-agent setting compounds this problem in ways that single-agent treatments do not anticipate. Most modern MARL systems, including multi-agent proximal policy optimization (MAPPO)\cite{yu2022surprising}, train a single policy network whose parameters are shared across all agents and updated from a batch that pools every agent's experience. Under this regime, a neuron that becomes dormant degrades the decision quality of every agent on the team rather than a single learner, so the cost of leaving plasticity loss untreated scales with team size. Diagnosis is also less straightforward: the same neuron may produce near-zero activation on one agent's input while remaining active on another's, and a single per-neuron statistic must summarize this heterogeneous behavior before any reset decision can be made. These properties of the shared-parameter regime: amplified damage, ambiguous diagnosis, and the composite nature of the training gradient discussed below, make plasticity recovery in cooperative MARL a problem that needs to be solved at the team level, not merely transplanted from single-agent recipes.

A natural intervention is to detect dormant neurons and reset their weights so that they can re-engage in learning~\cite{sokar2023dormant, dohare2024loss}. Carrying this out reliably in shared-parameter multi-agent training, however, raises two coupled challenges. The first is that existing methods declare a neuron dormant when its forward activation falls below a threshold~\cite{sokar2023dormant}. Yet under distributional shift, a neuron with near-zero activation may be in the process of being steered by gradient updates toward a new functional configuration, and resetting it on forward evidence alone destroys learning already underway~\cite{abbas2023loss}. The second is that adding a backward dimension, the natural refinement, is non-trivial: prior single-agent methods rely on hand-crafted proxies such as a utility score combining activation magnitude with outgoing weight norm~\cite{dohare2024loss}, or a gradient saliency value from an auxiliary forward-backward pass on the network output~\cite{he2025resin}. In MAPPO with shared parameters, the effective training gradient additionally reflects experience aggregation across agents, importance-ratio clipping, and advantage estimation, and whether hand-crafted surrogates remain reliable under this composite
signal has not previously been tested; we test it directly in
Section~\ref{sec:experiments}.

Existing plasticity maintenance methods have advanced unevenly along these two dimensions, and only a small subset has been examined under cooperative multi-agent training. Early countermeasures apply global perturbation indiscriminately, either by injecting noise into all parameters~\cite{10.5555/3495724.3496051} or by periodically resetting weights toward their initial values~\cite{nikishin2022primacy}, at the cost of disrupting features that remain useful. ReDo refines the granularity to individual neurons, resetting only those whose forward activation falls below a threshold~\cite{sokar2023dormant}, but inherits the limitation of single-dimensional detection. Notably, ReSiN~\cite{he2025resin} follows a similar neuron-level philosophy and adds a backward criterion, classifying a neuron as silent only when it is simultaneously forward-dormant and gradient-silent; it has been developed and evaluated only in single-agent settings, with the backward signal obtained through a separate forward--backward pass on the network output. On a parallel track, PE-MAMoE~\cite{qiu2026plasticity} addresses plasticity loss in cooperative multi-agent UAV emergency communication through an architectural route: a mixture-of-experts actor with a phase controller that perturbs experts after switches, preserving expressivity through expert diversity rather than neuron-level diagnosis and departing from the shared-MLP backbones standard in MARL practice.

Building on this analysis, we propose PRIME (\textbf{P}lasticity \textbf{R}ecovery \textbf{I}n \textbf{M}ulti-agent \textbf{E}nvironments), which extends bidirectional silent neuron detection to the shared-parameter multi-agent regime while remaining compatible with the standard MAPPO backbone~\cite{yu2022surprising}. The centralized-training-with-decentralized-execution (CTDE) architecture pools every agent's experience into a single set of network weights, providing a unified substrate for team-level diagnosis: PRIME aggregates forward activations and training gradients over the full team batch and classifies a neuron as \emph{silent} only when it is simultaneously forward-dormant and gradient-silent under layer-normalized thresholds. A forward-dormant neuron still receiving gradient updates is preserved.

The key distinction from prior bidirectional designs lies in the source of the backward signal. Rather than constructing a proxy gradient, PRIME reads the gradient that PPO training has already deposited on the network parameters, a signal that already encodes multi-agent data aggregation, importance-ratio clipping, and advantage computation. Detection is performed at fixed periodic intervals rather than triggered by external phase signals, decoupling plasticity recovery from any assumed knowledge of when the environment changes. When a silent neuron is reset, PRIME re-initializes its incoming weights while zeroing its outgoing connections, preserving policy output continuity across the reset.

The main contributions of this paper are as follows.
\begin{enumerate}
\item \textbf{Problem identification.}
  We provide what is, to our knowledge, the first systematic characterization of how dormant neuron accumulation limits adaptation across disaster phases in shared-parameter cooperative MARL. Diagnostic measurements show that forward dormancy and the actual training gradient can diverge substantially, particularly in the policy network, making a forward-only reset criterion prone to false positives. This internal-state perspective is distinct from prior multi-agent work based on architectural specialization.

\item \textbf{Method.}
  We propose PRIME, a plasticity maintenance module integrated into
shared-parameter MAPPO without altering its backbone architecture. Its core
contribution is a bidirectional silent neuron criterion built on three
deliberate choices: the backward signal is the gradient the PPO training
loss has already deposited, so diagnosis reflects each neuron's actual role
in the composite multi-agent objective (experience aggregation,
importance-ratio clipping, advantage estimation); detection statistics
aggregate over the full team batch rather than any single agent's slice;
and the trigger is a fixed internal period rather than an external phase
signal. To our knowledge, this is the first plasticity maintenance method
designed for the shared-parameter cooperative MARL regime that performs
neuron-level diagnosis on the live training gradient.

\item \textbf{Theoretical and empirical validation.}
  Extending the single-agent tracking analysis of ReSiN~\cite{he2025resin} to
the team-averaged MAPPO objective, we establish a dynamic regret bound
showing that selectively resetting only the identified silent neurons incurs
lower tracking cost than perturbing all parameters. Experiments in a
non-stationary UAV-ECN environment with eight phase transitions show a
24.9\% IQM improvement over MAPPO and suppression of dormant neuron
fractions from 40--45\% to 10--20\%, with consistent gains in coverage and
collision safety; design-axis ablations that substitute the backward signal,
the aggregation slice, and the reset operator in isolation confirm that the
first two choices carry the effect while the operator is exchangeable.
\end{enumerate}

The remainder of this paper is organized as follows. Section~\ref{sec:motivation} presents the measurements that motivate the design. Section~\ref{sec:system} describes the system model and problem formulation. Section~\ref{sec:method} details the PRIME framework, including the bidirectional detection mechanism, the reset procedure, and the theoretical analysis. Section~\ref{sec:experiments} reports experimental results and ablation studies. Section~\ref{sec:related} reviews related work, and Section~\ref{sec:conclusion} concludes the paper.

\section{Motivation and Analysis}
\label{sec:motivation}

This section establishes three findings on a vanilla MAPPO controller in our UAV-ECN simulator: dormant neurons accumulate and persist under standard gradient updates; forward dormancy and backward gradient silence diverge in three of four hidden layers, so a forward-only reset criterion would discard neurons that are still being actively updated; and dormancy decisions vary by $5$--$18\%$ across agents, making per-agent reset criteria structurally inconsistent. Together these findings define the design target for Section~\ref{sec:method}.

\subsection{Plasticity Loss Accumulates and Persists}
\label{subsec:plasticity_loss}

In a UAV-ECN mission the optimization objective changes at every phase boundary (Section~\ref{sec:system} details the phase model), violating the stationary-MDP assumption underlying standard PPO convergence guarantees~\cite{schulman2017proximal} and creating conditions under which plasticity loss is both rapid and compounding.

To quantify the effect, we instrument a vanilla MAPPO controller (shared two-layer MLP, width $H{=}32$, ReLU activations) under two regimes: a fixed-phase \textit{Normal} mode ($3.07{\times}10^6$ environment steps) and a cyclic-phase-switching \textit{Change} mode with eight phase transitions ($18.43{\times}10^6$ steps). At every training iteration we record (i)~the \emph{dormant neuron fraction} (DNF), defined as the proportion of hidden neurons whose batch-averaged absolute activation falls below a normalized threshold $\tau_d$; and (ii)~the \emph{dormant overlap} (LDO), the fraction of all hidden neurons that are dormant at both the current and previous checkpoint.

\begin{figure}[!t]
\centering
\includegraphics[width=0.9\columnwidth]{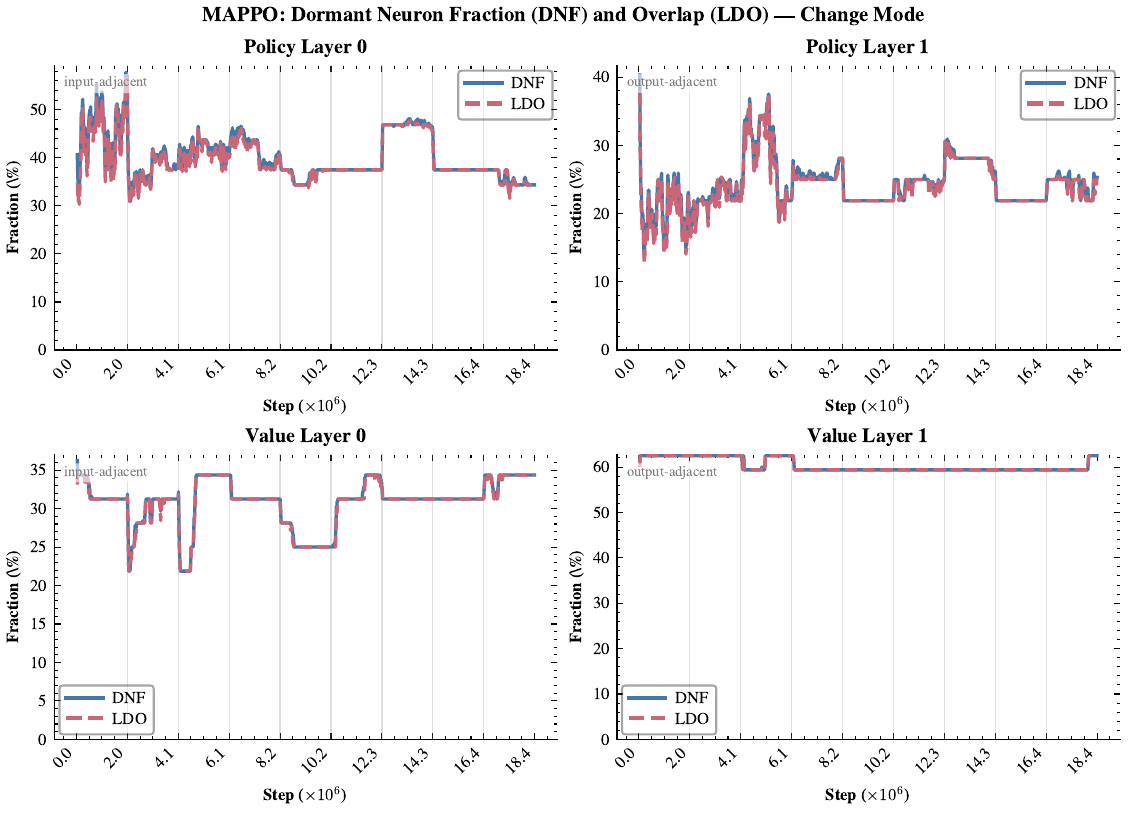}
\caption{Dormant Neuron Fraction (DNF, solid blue) and Dormant Overlap (LDO, dashed red) for the four hidden layers of a vanilla MAPPO controller in Change mode ($\tau_d{=}0.5$; vertical gray lines mark the eight phase boundaries, none of which produces spontaneous recovery).
The near-coincidence of LDO with DNF indicates that the current dormant set is almost entirely a subset of the previous one rather than rotating across iterations.
Value Layer~1 saturates near $60\%$; Policy Layer~0 ranges $31$--$58\%$, Policy Layer~1 $14$--$40\%$, and Value Layer~0 $22$--$36\%$.}
\label{fig:motivation}
\end{figure}

Figure~\ref{fig:motivation} reveals a non-uniform but broadly severe dormancy landscape. Normal-mode counterparts of this and the two following figures are collected in Appendix~D of the supplementary material. Value Layer~1 saturates near $60\%$ in both modes, leaving a majority of that layer effectively unused, and the remaining three layers fluctuate in $14$--$58\%$ bands without spontaneously decaying. Across all four layers, DNF and LDO traces are nearly indistinguishable, meaning that the current dormant set is almost entirely a subset of the previous one: dormancy persists across checkpoints rather than rotating. In Change mode, the eight phase boundaries produce no spontaneous recovery; DNF in the affected layers oscillates around its phase-conditional level but never relaxes. Standard gradient updates alone are therefore insufficient to reactivate dormant units, and the cooperative CTDE architecture compounds the cost: because all $n$ agents share the same network, every dormant neuron degrades every agent's output simultaneously.

\subsection{Forward Dormancy Diverges from Backward Gradient Silence}
\label{subsec:decoupling}

Forward DNF measures whether a neuron contributes to the network's output at inference time, but it does not measure whether the optimizer is still investing learning capacity in that neuron. To make the distinction concrete, we additionally measure the \emph{backward silence fraction} $|\mathcal{G}_l|/H_l$, where $\mathcal{G}_l$ is the set of neurons whose per-parameter gradient magnitude, normalized within the layer, falls below a threshold $\tau_g$. Both quantities come from the same run under the same per-layer normalization ($\tau_d{=}0.5$, $\tau_g{=}0.08$), so the two fractions are directly comparable.

\begin{figure}[!t]
\centering
\includegraphics[width=0.9\columnwidth]{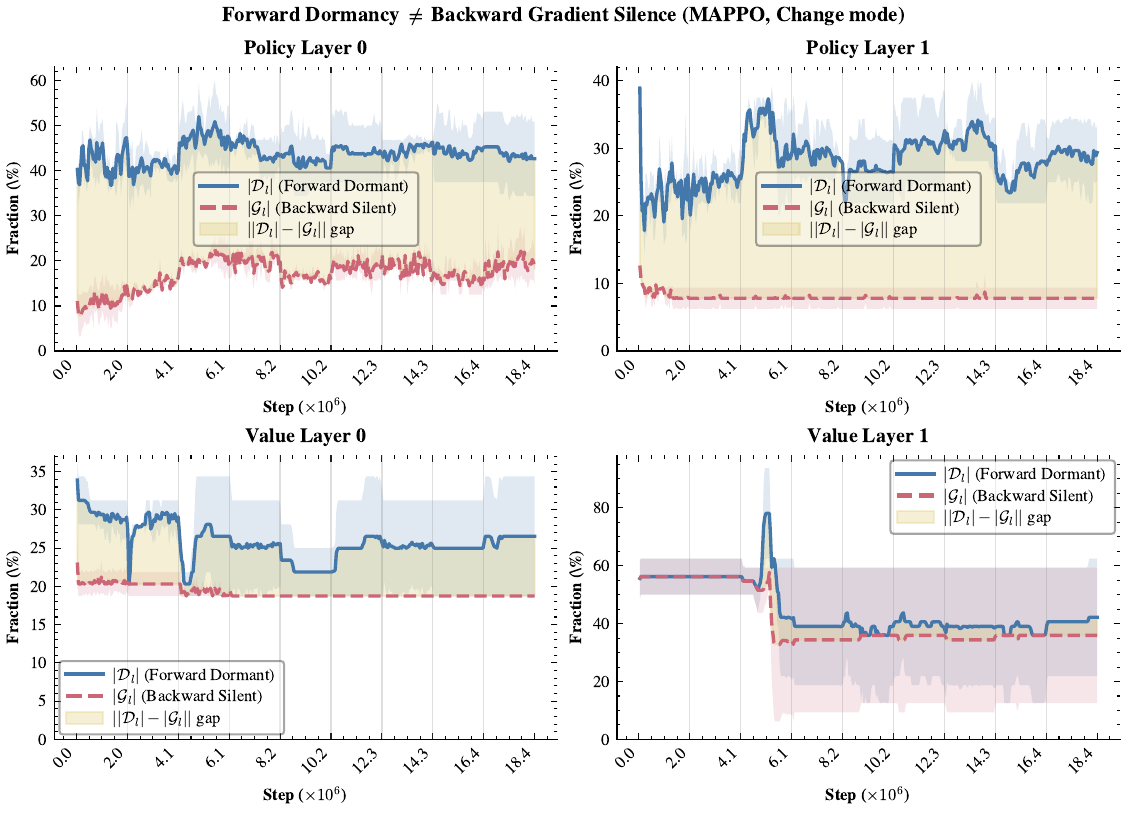}
\caption{Forward-dormant fraction $|\mathcal{D}_l|/H_l$ (solid blue, $\tau_d{=}0.5$) and backward-silent fraction $|\mathcal{G}_l|/H_l$ (dashed red, $\tau_g{=}0.08$) on a vanilla MAPPO controller in Change mode; both quantities use the same per-layer normalization.
The shaded region $||\mathcal{D}_l|-|\mathcal{G}_l||$ marks the magnitude of decoupling.
Policy Layers~0--1 and Value Layer~0 show a persistent gap with $|\mathcal{D}_l| > |\mathcal{G}_l|$, while Value Layer~1 saturates with $|\mathcal{D}_l| \approx |\mathcal{G}_l|$.}
\label{fig:decoupling}
\end{figure}

Figure~\ref{fig:decoupling} shows the resulting time series. The forward-dormant set is strictly larger than the backward-silent set in three of four hidden layers throughout training. In Policy Layer~0 the gap reaches $20$--$30$ percentage points ($|\mathcal{D}_l|{\approx}45\%$ vs $|\mathcal{G}_l|{\approx}17\%$); in Policy Layer~1, $|\mathcal{D}_l|{\approx}30\%$ versus $|\mathcal{G}_l|{\approx}10\%$; in Value Layer~0, $|\mathcal{D}_l|{\approx}33\%$ versus $|\mathcal{G}_l|{\approx}18\%$. Value Layer~1 is the only exception: $|\mathcal{D}_l|$ and $|\mathcal{G}_l|$ both saturate at ${\approx}62\%$ and remain numerically indistinguishable for the entire run, reflecting a fully degenerate state in which the two criteria converge on the same neurons.

\begin{figure}[!t]
\centering
\includegraphics[width=0.9\columnwidth]{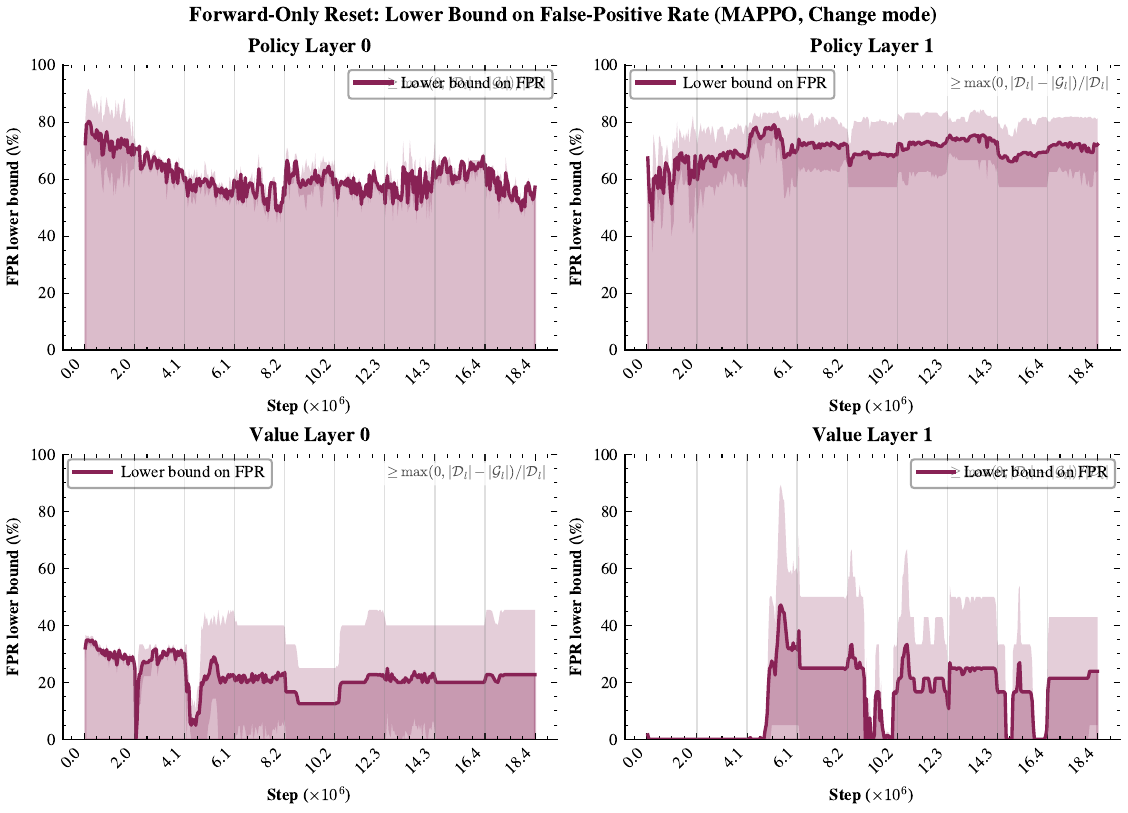}
\caption{Lower bound on the false-positive rate of a forward-only reset, $\max(0,|\mathcal{D}_l|-|\mathcal{G}_l|)/|\mathcal{D}_l|$, on a vanilla MAPPO controller in Change mode.
The bound under-estimates the rate at which a forward-only criterion would flag neurons that still receive non-trivial gradient signal; Value Layer~1 remains near zero because dormancy and gradient silence have converged at saturation.}
\label{fig:fpr_lb}
\end{figure}

The persistent gap in the three non-saturated layers carries a concrete consequence for reset design: the set difference $\mathcal{D}_l \setminus \mathcal{G}_l$ consists of neurons whose forward activation is near zero but whose gradients remain large, neurons the optimizer is still actively steering toward the incoming distribution. A reset criterion that flags neurons solely on the basis of forward dormancy, as in ReDo~\cite{sokar2023dormant}, would discard the optimizer's accumulated progress on this population. Figure~\ref{fig:fpr_lb} bounds this cost from below by $\max(0,|\mathcal{D}_l|-|\mathcal{G}_l|)/|\mathcal{D}_l|$, an under-estimate of the false-positive rate that a forward-only reset would incur. In Policy Layer~0 this bound reaches $50$--$90\%$; in Policy Layer~1 it stays in $55$--$75\%$; in Value Layer~0 it ranges $25$--$45\%$; the true false-positive rate is likely higher.

A criterion that intersects $\mathcal{D}_l$ with $\mathcal{G}_l$ resolves this by definition, preserving forward-dormant neurons that still receive gradient signal. Value Layer~1 illustrates the complementary case: at saturation the intersection gracefully reduces to the forward-only criterion ($\mathcal{D}_l \cap \mathcal{G}_l = \mathcal{D}_l$), so the combined criterion is strictly safer than either component alone.

\subsection{Cross-Agent Dormancy Disagreement}
\label{subsec:cross_agent}

Shared-parameter CTDE routes $n$ agents' observations through a single network, but each agent sees the joint state from its own perspective, so whether a neuron is dormant can depend on whose observation is being processed.

\begin{figure}[!t]
\centering
\includegraphics[width=0.9\columnwidth]{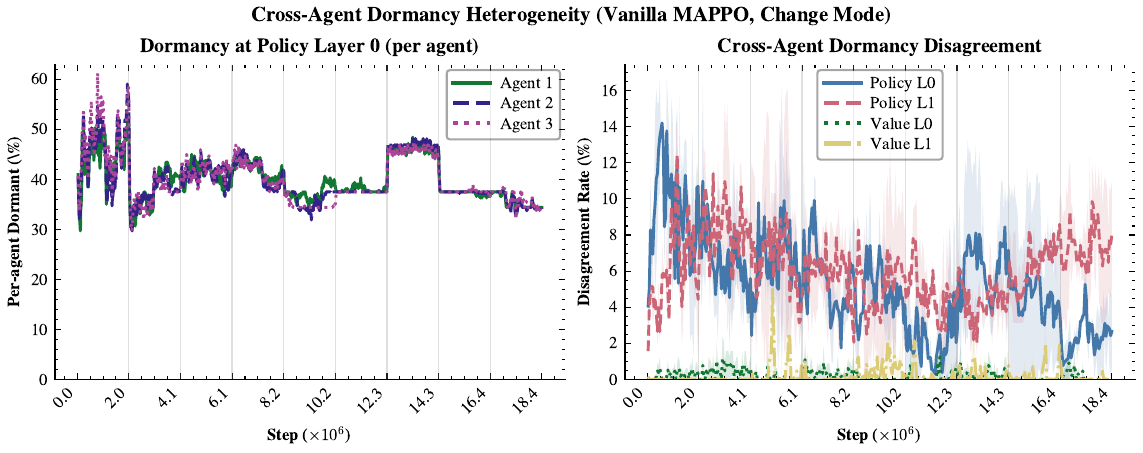}
\caption{Cross-agent dormancy heterogeneity on a vanilla MAPPO controller in Change mode.
(a)~Per-agent forward dormancy at Policy Layer~0: the three agents trace visibly distinct trajectories within a $35$--$55\%$ band, indicating that the network's effective dormancy varies with input source.
(b)~Cross-agent disagreement rate, $(|\bigcup_i \mathcal{D}_l^{(i)}| - |\bigcap_i \mathcal{D}_l^{(i)}|)/H_l$, across all four layers.
Policy layers exhibit $5$--$18\%$ persistent disagreement; value layers remain near zero because the centralized critic operates on a joint-state encoding.}
\label{fig:cross_agent}
\end{figure}

Figure~\ref{fig:cross_agent}(a) shows per-agent forward dormancy at Policy Layer~0 across training. The three agents trace visibly distinct trajectories within a $35$--$55\%$ band. Figure~\ref{fig:cross_agent}(b) quantifies this dependence through the disagreement rate, defined as $(|\bigcup_i \mathcal{D}_l^{(i)}| - |\bigcap_i \mathcal{D}_l^{(i)}|)/H_l$, where $\mathcal{D}_l^{(i)}$ is the dormant set under agent $i$'s observations alone. Disagreement is modest but persistent at $5$--$18\%$ in policy layers; in value layers it remains near zero because the centralized critic operates on a joint-state encoding common to all agents.

A per-agent reset criterion therefore yields inconsistent decisions at every checkpoint; resolving the conflicts requires either a voting scheme, which discards information, or a unified statistic over the full $B{\times}n$ team batch. PRIME adopts the latter (Section~\ref{subsec:detection}).

\section{System Model}
\label{sec:system}

\begin{figure*}[!t]
\centering
\includegraphics[width=0.9\textwidth]{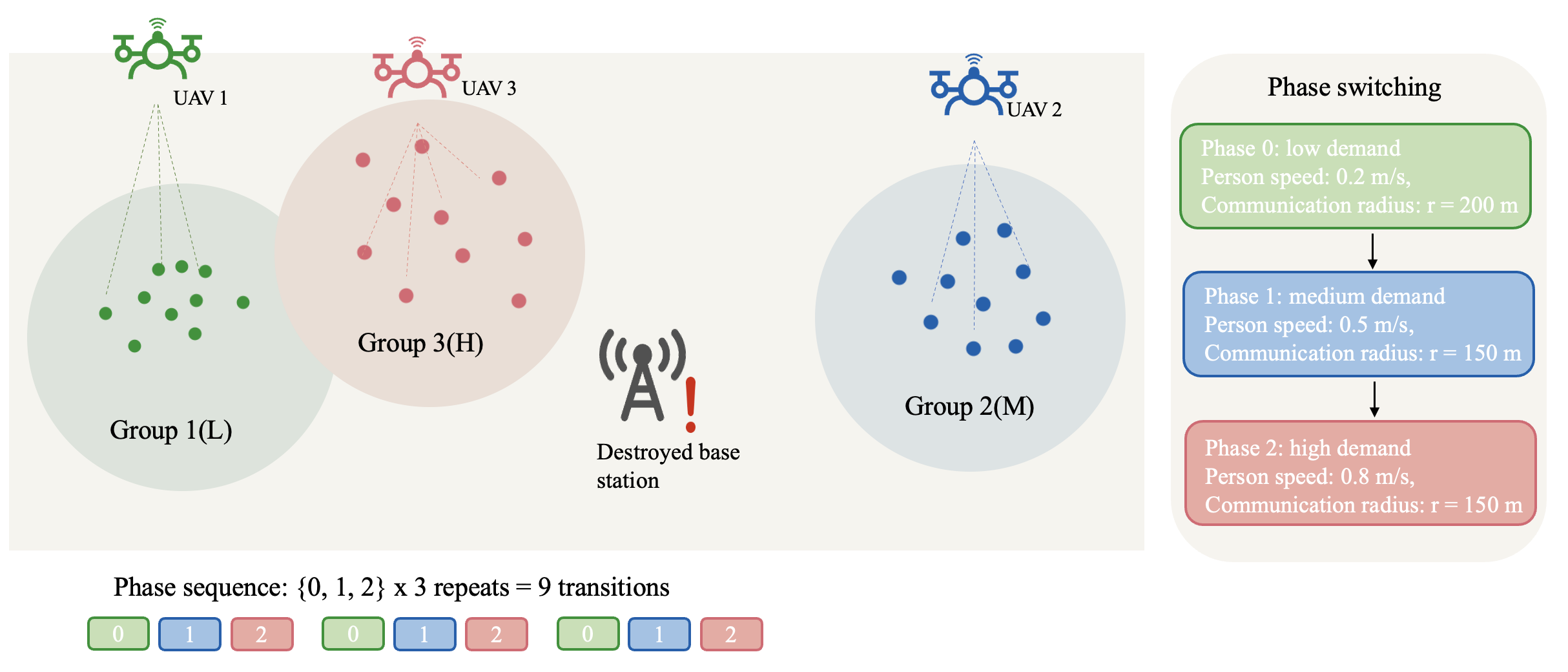}
\caption{System model of the UAV-assisted emergency communication network.
Three UAVs serve as aerial base stations for $N{=}20$ ground users partitioned into $G{=}3$ RPGM groups under a cyclic three-phase demand regime.
Each phase alters user mobility speed, communication coverage radius, and demand class (L/M/H).
The phase sequence $\{0,1,2\}$ repeats three times, yielding nine phase segments (eight phase transitions) that drive the non-stationarity analyzed in this work.}
\label{fig:system_model}
\end{figure*}

Following the UAV-ECN setup of~\cite{qiu2026plasticity}, this section introduces the system model and its Dec-POMDP abstraction; Fig.~\ref{fig:system_model} illustrates the scenario.

\subsection{Network Topology and Kinematics}

Consider a post-disaster square area $\Omega = [0,S]^2$ in which terrestrial infrastructure has been destroyed (Fig.~\ref{fig:system_model}).
A fleet of $U$~rotary-wing UAVs, indexed by $\mathcal{U} = \{1,\ldots,U\}$, is deployed as aerial base stations to provide temporary communication coverage for $N$ ground users.
At discrete time slot~$t$, the three-dimensional position and velocity of UAV~$u$ are
\begin{equation}
  \mathbf{q}_u(t) = \bigl[x_u(t),\, y_u(t),\, h_u(t)\bigr]^\top,
  \mathbf{v}_u(t) = \bigl[v_{h,u}(t),\, v_{z,u}(t)\bigr]^\top,
\end{equation}
with the kinematic update $\mathbf{q}_u(t{+}1) = \mathbf{q}_u(t) + \mathbf{v}_u(t)\,\Delta t$,
subject to maximum horizontal speed $V_{\max}^h$, maximum vertical speed $V_{\max}^z$, altitude bounds $[h_{\min}, h_{\max}]$, and the area boundary $\mathbf{q}_u(t) \in \Omega$.

\subsection{Channel and Interference Model}

The line-of-sight (LoS) probability for UAV--user pair $(u,n)$ follows the 3GPP Urban Micro model with altitude correction~\cite{3gpp2020}:
\begin{equation}
P_{\mathrm{LoS}}(r,h) = \frac{18}{r} + e^{-r/36}\!\left(1-\frac{18}{r}\right)\!\bigl(1+C_0(h)\bigr),
\end{equation}
where $C_0(h)=\max\{h{-}13,\,0\}/101.5$ and $r$ is the horizontal distance.
This closed-form approximation of the 3GPP TR~36.777 LoS table reduces simulator overhead and is adopted from~\cite{qiu2026plasticity}.
Path loss follows the 3GPP TR~38.901 large-scale models with Rician fading under LoS and Nakagami-$m$ fading under NLoS.
A frequency reuse factor $F_{\mathrm{reuse}}$ partitions UAVs into co-channel groups.
The aggregate interference experienced by user~$n$ when served by UAV~$u$ is
\begin{equation}
I_{n,u}(t) = \!\!\sum_{\substack{u' \neq u \\ c(u'){=}c(u)}}\!\! P_{u'}g_{n,u'}
  + \rho_{\mathrm{adj}}\!\!\sum_{\substack{u' \\ c(u'){\neq}c(u)}}\!\! P_{u'}g_{n,u'},
\end{equation}
where $\rho_{\mathrm{adj}} \in [0,1)$ is the adjacent-channel leakage ratio, $P_{u'}$ is the transmit power, and $g_{n,u'}$ is the composite channel gain including path loss and small-scale fading.

\subsection{Energy Consumption Model}

Per-slot propulsion energy follows the three-dimensional rotary-wing model of~\cite{yan2021new,zeng2019energy}:
\begin{equation}
\begin{aligned}
P_{\mathrm{rot}}(V_h, V_z) = 
& P_0\!\left(1 + \frac{3V_h^2}{U_{\mathrm{tip}}^2}\right) \\
& + P_i\!\left(\!\sqrt{1+\frac{(V_h^2{+}V_z^2)^2}{4v_0^4}}-\frac{V_h^2{+}V_z^2}{2v_0^2}\right)^{\!\!1/2}\!\! \\
& + \tfrac{1}{2}d_0\rho s A V_h^3,
\end{aligned}
\end{equation}
with standard rotary-wing aerodynamic parameters (blade profile and induced power $P_0$, $P_i$; tip speed $U_{\mathrm{tip}}$; hover induced velocity $v_0$; drag ratio $d_0$; air density $\rho$; solidity $s$; disc area $A$).
This energy cost enters the reward function as a penalty term that the policy must balance against coverage objectives.

\subsection{User Mobility and Phase Model}

$N$~ground users are distributed among $G$~hotspots according to the Reference Point Group Mobility (RPGM) model~\cite{camp2002survey,hong1999group}.
Individual velocities combine a Gauss--Markov process with a group-following attraction:
\begin{equation}
\mathbf{v}_n^+(t) = (1{-}\kappa)\,\mathbf{v}_n^{\mathrm{GM}}(t) + \kappa v_0 \frac{\mathbf{c}_g(t)-\mathbf{x}_n(t)}{\|\mathbf{c}_g(t)-\mathbf{x}_n(t)\|+\varepsilon} + \boldsymbol{\eta}_n(t),
\end{equation}
where $\kappa \in [0,1]$ controls the group-cohesion strength.

Time is partitioned into phases $\varphi_t \in \{0,1,2\}$, each of duration $L$~iterations, cycling as $\varphi_t = \lfloor t/L \rfloor \bmod 3$.
As shown in the right panel of Fig.~\ref{fig:system_model}, each phase defines a distinct mobility--demand regime:
Phase~0 (low demand, user speed $0.2$, coverage radius $r{=}200\,$m),
Phase~1 (medium demand, speed $0.5$, $r{=}150\,$m), and
Phase~2 (high demand, speed $0.8$, $r{=}150\,$m).
The demand level of each user group is drawn from classes $\ell \in \{\text{L},\text{M},\text{H}\}$ with corresponding rate thresholds and reward weights (Table~\ref{tab:env_params}).
The sequence $\{0,1,2\}$ repeats three times, yielding nine phase segments and eight phase transitions that constitute the structured non-stationarity of the UAV-ECN task.

\subsection{Multi-Objective Optimization}

At each slot~$t$, the controller jointly selects UAV poses, binary user associations $a_{n,u} \in \{0,1\}$, and per-link bandwidth/power allocations to maximize
\begin{equation}
\begin{aligned}
\max_{\mathbf{q},\mathbf{v},\mathbf{a},\mathbf{b},\mathbf{p}} \;
& \mu_t^{\mathrm{qoe}}\!\sum_{n,u}\! a_{n,u}\, w_{\ell_n}\!\bigl[R_{n,u} \!\geq\! r_{\ell_n}\bigr]
-\mu_t^{\mathrm{ene}}\!\sum_u\! E_u^{\mathrm{prop}} \\
& -\mu_t^{\mathrm{col}}\!\sum_{u<u'}\!\bigl[\|\mathbf{q}_u{-}\mathbf{q}_{u'}\| \!<\! d_{\min}\bigr],
\end{aligned}
\label{eq:opt}
\end{equation}
subject to battery dynamics, association and power-budget constraints, and minimum inter-UAV separation $d_{\min}$~\cite{qiu2026plasticity}.
The phase-varying weights $(\mu_t^{\mathrm{qoe}}, \mu_t^{\mathrm{ene}}, \mu_t^{\mathrm{col}})$ shift at every phase boundary, jointly defining a piecewise-stationary objective whose transitions drive the plasticity dynamics this paper addresses.

\subsection{Dec-POMDP Formulation}

We cast Problem~\eqref{eq:opt} as a cooperative decentralized partially observable Markov decision process (Dec-POMDP)~\cite{amato2024introduction}, defined by the tuple $\langle n, \mathcal{S}, \{\mathcal{A}_u\}_{u=1}^n, \{\mathcal{O}_u\}_{u=1}^n, P, R, \gamma\rangle$, where $n$ is the team size, $\mathcal{S}$ is the joint state space encompassing UAV poses, user positions, channel realizations, and the current phase index $\varphi_t$, and $P$ is the joint transition kernel induced by UAV kinematics, user mobility, and channel dynamics.
At each slot $t$, UAV~$u$ receives a local observation $o_u(t) \in \mathcal{O}_u$ comprising its own 3D position, relative positions of teammates, residual battery level, positions and achieved rates of users within its service radius, and the current-phase demand weights.
Its action $a_u(t) \in \mathcal{A}_u$ selects the next pose displacement and the associations and resources for users it currently serves.
The team shares a scalar reward $R$ equal to the objective in~\eqref{eq:opt}, normalized per step, and discounts with $\gamma \in [0,1)$.

We adopt the centralized training with decentralized execution (CTDE) paradigm~\cite{lowe2017multi}: during training, a centralized critic conditioned on the joint state evaluates a shared actor that conditions on local observations only, while at execution each UAV acts independently using its local policy.
All $n$~UAVs share the same actor parameters, which both reduces sample complexity and ensures that the dormancy and gradient statistics aggregated at the team level (Section~\ref{subsec:detection}) are well-posed.

\section{Methodology: PRIME}
\label{sec:method}

PRIME augments the standard MAPPO training loop with a periodic neuron activity assessment and selective reinitialization step.
The network architecture remains unchanged: a shared MLP actor with two hidden layers of width $H$ and ReLU activations, paired with a centralized MLP critic of identical topology.
No routing modules, expert sub-networks, or auxiliary regularization terms are added.
The only modification is a diagnostic-and-reset procedure executed every $F$ mini-batch steps on both the actor and the critic.

The design is guided by a tradeoff between two failure modes observed in Section~\ref{sec:motivation}.
A reset criterion that is too permissive (such as the forward-only rule of ReDo~\cite{sokar2023dormant}) reinitializes neurons that still receive informative gradient signal, incurring substantial false positives (Section~\ref{subsec:decoupling}).
A reset criterion that is too conservative leaves dormant neurons in place, allowing plasticity to erode across phases.
PRIME navigates between these regimes by intersecting two complementary signals: forward dormancy identifies neurons whose information contribution has collapsed, while backward silence verifies that the optimizer has indeed abandoned them.

A second design consideration, the $5$--$18\%$ per-agent dormancy disagreement measured in Section~\ref{subsec:cross_agent}, is addressed by aggregating both detection statistics over the full $B{\times}n$ team batch (Section~\ref{subsec:detection}).

\subsection{Preliminaries: Dormant and Silent Neurons}
\label{subsec:prelim}

Under the phase-driven non-stationarity of Section~\ref{sec:system}, the Dec-POMDP becomes a family $\{M_0, M_1, M_2\}$ whose kernels $P_\varphi$ and rewards $R_\varphi$ are indexed by the demand regime; the mission cycles $\varphi_t \in \{0,1,2\}$ three times, producing eight transitions.
The shared policy $\pi_\theta$ must therefore track a piecewise-stationary optimum $w_\varphi^*$ that jumps at each transition, motivating the path-length-based tracking analysis in Section~\ref{subsec:theory}.

Within each $M_\varphi$, let $f_\theta\colon \mathbb{R}^{d_{\mathrm{in}}} \!\to\! \mathbb{R}^{d_{\mathrm{out}}}$ denote the shared policy network.
Layer~$l$ contains $H_l$ neurons with post-activation outputs $h_{l,i}(\mathbf{x}) = \sigma(\mathbf{w}_{l,i}^\top \mathbf{x} + b_{l,i})$.

\begin{definition}[Dormant Neuron~\cite{sokar2023dormant}]
\label{def:dormant}
The \emph{dormancy index} of neuron $(l,i)$ over data distribution $\mathcal{D}$ is the ratio of its mean absolute activation to the layer average:
\begin{equation}
s_{l,i} \;=\; \frac{\mathbb{E}_{\mathbf{x}\sim\mathcal{D}}\bigl|h_{l,i}(\mathbf{x})\bigr|}
{\frac{1}{H_l}\sum_{j=1}^{H_l}\mathbb{E}_{\mathbf{x}\sim\mathcal{D}}\bigl|h_{l,j}(\mathbf{x})\bigr|}.
\end{equation}
Neuron $(l,i)$ is \emph{dormant} if $s_{l,i} \leq \tau_d$ for a user-specified threshold $\tau_d > 0$.
\end{definition}

\begin{definition}[Silent Neuron~\cite{he2025resin}]
\label{def:silent}
The \emph{activity index} of neuron $(l,i)$ combines its forward activation with its backward gradient:
\begin{equation}
\xi_{l,i} \;=\; \frac{\mathbb{E}_{\mathbf{x}\sim\mathcal{D}}\bigl|h_{l,i}(\mathbf{x})\bigr| \;+\; \mathbb{E}_{\mathbf{x}\sim\mathcal{D}}\bigl|g_{l,i}(\mathbf{x})\bigr|}
{\frac{1}{H_l}\sum_{j=1}^{H_l}\mathbb{E}_{\mathbf{x}\sim\mathcal{D}}\bigl|h_{l,j}(\mathbf{x})\bigr|},
\end{equation}
where $g_{l,i}(\mathbf{x}) = \partial \sum_n f_\theta(\mathbf{x}_n)/\partial h_{l,i}$ is the gradient of the aggregated network output with respect to the neuron's activation.
A neuron is \emph{silent} when $\xi_{l,i} < \varepsilon$, which, under the boundedness and non-degeneracy conditions of~\cite{he2025resin}, holds if and only if both $\mathbb{E}|h_{l,i}|$ and $\mathbb{E}|g_{l,i}|$ are vanishingly small.
\end{definition}

The key distinction between dormancy and silence is that a dormant neuron may still receive substantial gradient signal, indicating that the optimizer is actively steering it toward a useful configuration.

\subsection{Multi-Agent Silent Neuron Detection}
\label{subsec:detection}

PRIME adapts the single-agent definitions above to the batch structure of shared-parameter CTDE training.
Each rollout of length $T_{\mathrm{roll}}$ collects observations from all $n$~UAVs, producing a flattened batch $\mathbf{O}$ of shape $(T_{\mathrm{roll}} \!\cdot\! n,\; d_{\mathrm{obs}})$.
The shared actor processes this batch in a single forward pass, yielding an activation tensor of shape $(T_{\mathrm{roll}} \!\cdot\! n,\; H_l)$ at each hidden layer~$l$.

\textbf{Forward activity index.}
Per-neuron activations $\{h_{l,i}^{(k)}\}_{k=1}^{B \cdot n}$ recorded during the PPO update are aggregated into a normalized dormancy score:
\begin{equation}
\hat{s}_{l,i} \;=\; \frac{\frac{1}{Bn}\sum_{k} |h_{l,i}^{(k)}|}
{\frac{1}{H_l}\sum_{j}\frac{1}{Bn}\sum_{k} |h_{l,j}^{(k)}|}.
\label{eq:forward_index}
\end{equation}
Neuron $(l,i)$ is flagged as \emph{forward-dormant} when $\hat{s}_{l,i} \leq \tau_d$.
Because the batch spans all $n$~agents, this score measures team-level dormancy rather than any single agent's observation distribution.

\textbf{Backward activity index.}
During the $K_{\mathrm{epoch}}$ PPO update epochs, per-neuron weight gradients $\partial\mathcal{L}_{\mathrm{PPO}}/\partial\mathbf{w}_{l,i}$ are accumulated across all mini-batches.
The normalized gradient magnitude is
\begin{equation}
\hat{g}_{l,i} \;=\; \left\|\frac{\partial\mathcal{L}_{\mathrm{PPO}}}{\partial\mathbf{w}_{l,i}}\right\|_1 \!\Big/\, |\mathbf{w}_{l,i}|,
\label{eq:backward_index}
\end{equation}
where $|\mathbf{w}_{l,i}|$ denotes the number of elements in the weight vector of neuron~$i$.
Neuron $(l,i)$ is flagged as \emph{backward-silent} when $\hat{g}_{l,i} \leq \tau_g$.

\textbf{Intersection criterion.}
Let $\mathcal{D}_l = \{i \colon \hat{s}_{l,i} \leq \tau_d\}$ denote the forward-dormant set and $\mathcal{G}_l = \{i \colon \hat{g}_{l,i} \leq \tau_g\}$ the backward-silent set.
The \emph{silent neuron set} is their intersection:
\begin{equation}
\mathcal{S}_l \;=\; \mathcal{D}_l \;\cap\; \mathcal{G}_l.
\label{eq:silent_set}
\end{equation}
Only neurons in $\mathcal{S}_l$ are candidates for reinitialization.
No neuron carrying an informative signal in \emph{either} the forward or the backward direction is disturbed.

Two properties of this construction deserve note. The backward index costs
nothing beyond training itself: the gradient in~\eqref{eq:backward_index} is
the one the optimizer has already computed, at the scale the optimizer
actually operates on, so no auxiliary pass is required. A known concern with
loss-based gradients is masking: a neuron may receive a vanishing loss
gradient while remaining active in the forward pass~\cite{he2025resin}. The
intersection in~\eqref{eq:silent_set} neutralizes this failure mode, since
such a neuron fails the forward-dormancy test and is never reset.
Section~\ref{sec:experiments} tests the converse design directly, replacing
the training gradient with an output-sensitivity proxy.

\subsection{Selective Reinitialization}
\label{subsec:reinit}

For each silent neuron $(l,i) \in \mathcal{S}_l$, three coordinated operations restore its capacity while preventing immediate disruption of downstream layers:
\begin{align}
\mathbf{w}_{l,i} &\leftarrow \mathrm{Uniform}\!\bigl(-b_l,\, b_l\bigr)^{d_{l,\mathrm{in}}},\quad b_l = \sqrt{3/d_{l,\mathrm{in}}},\label{eq:reset_in}\\
b_{l,i} &\leftarrow 0,\label{eq:reset_bias}\\
[\mathbf{W}_{l+1}]_{:,i} &\leftarrow \mathbf{0}.\label{eq:reset_out}
\end{align}
Step~\eqref{eq:reset_in} draws the incoming weights from the Kaiming uniform distribution, matching the variance of the original initialization.
Step~\eqref{eq:reset_bias} zeros the bias.
Step~\eqref{eq:reset_out} sets the corresponding column of the next layer's weight matrix to zero, so the freshly reinitialized neuron produces no immediate output perturbation and must be activated through subsequent gradient updates~\cite{sokar2023dormant}.
The Adam optimizer state (first and second moment estimates) for all affected parameter slices is also cleared to prevent stale momentum from biasing the relearning trajectory.
The operator ablation in Section~\ref{sec:experiments} indicates that the
hard reset itself is not performance-critical: a matched-cadence stochastic
perturbation of the same silent set performs on par, and the reset is
retained for its output continuity at the moment of intervention.

The same reset procedure is applied independently to the shared actor and the centralized critic.

\subsection{PRIME-MAPPO Training Procedure}
\label{subsec:algorithm}

Algorithm~\ref{alg:prime} summarizes the procedure. Each iteration
collects a multi-agent rollout, flattens it into the
$(T_{\mathrm{roll}} \!\cdot\! n,\; d_{\mathrm{obs}})$ batch of
Section~\ref{subsec:detection}, and runs standard clipped-PPO updates with
GAE~\cite{schulman2016high}; the only additions are the per-neuron
activation and gradient statistics recorded during each forward and
backward pass, at a cost of two scalars per neuron per layer. Every $F$
mini-batch steps, the detection of Section~\ref{subsec:detection} and the
reset of Section~\ref{subsec:reinit} are applied to both the actor and the
critic; with $F{=}200$ and $K_{\mathrm{epoch}} \times 32 = 256$ mini-batch
steps per iteration, detection fires approximately once per training
iteration.

\begin{algorithm}[!t]
\small
\caption{PRIME-MAPPO Training}
\label{alg:prime}
\begin{algorithmic}[1]
\Require Shared actor $\pi_\theta$, centralized critic $V_\psi$ (both two-layer MLP, width $H$, ReLU); thresholds $\tau_d$, $\tau_g$; reset period $F$ (in mini-batch steps); $n$~agents; PPO hyperparameters ($K_{\mathrm{epoch}}$, clip $\epsilon$, $\gamma$, $\lambda_{\mathrm{GAE}}$)
\Ensure Trained policy $\pi_\theta$
\State Initialize $\theta$, $\psi$ via Kaiming uniform; set step counter $c \leftarrow 0$
\For{iteration $t = 1, 2, \ldots, T$}
  \State \textbf{// Step 1: Multi-agent rollout}
  \For{step $k = 1, \ldots, T_{\mathrm{roll}}$}
    \For{each UAV $u = 1, \ldots, n$ \textbf{in parallel}}
      \State Observe $o_u^{(k)}$; sample $a_u^{(k)} \sim \pi_\theta(\cdot \mid o_u^{(k)})$
    \EndFor
    \State Execute joint action; receive shared reward $r^{(k)}$
  \EndFor
  \State Flatten: $\mathbf{O} \leftarrow \{o_u^{(k)}\}_{u,k}$, shape $(T_{\mathrm{roll}} \!\cdot\! n,\; d_{\mathrm{obs}})$
  \State \textbf{// Step 2: MAPPO update with activity recording}
  \State Compute GAE advantages $\hat{A}$ using $V_\psi$
  \For{epoch $e = 1, \ldots, K_{\mathrm{epoch}}$}
    \For{each mini-batch $\mathcal{B} \subset \mathbf{O}$}
      \State Forward pass: record activations $\{h_{l,i}^{(k)}\}$
      \State Compute clipped PPO loss $\mathcal{L}_{\mathrm{PPO}}$ and value loss $\mathcal{L}_V$
      \State Backpropagate; update $\theta$, $\psi$ via Adam; $c \leftarrow c + 1$
      \State \textbf{// Step 3: Periodic silent neuron detection \& reset}
      \If{$c \bmod F = 0$}
        \For{each hidden layer $l$ in $\pi_\theta$ and $V_\psi$}
          \State Compute $\hat{s}_{l,i}$ via~\eqref{eq:forward_index}, $\hat{g}_{l,i}$ via~\eqref{eq:backward_index}
          \State $\mathcal{D}_l \!\leftarrow\! \{i : \hat{s}_{l,i} \!\leq\! \tau_d\}$;\; $\mathcal{G}_l \!\leftarrow\! \{i : \hat{g}_{l,i} \!\leq\! \tau_g\}$;\; $\mathcal{S}_l \!\leftarrow\! \mathcal{D}_l \!\cap\! \mathcal{G}_l$
          \For{each $i \in \mathcal{S}_l$}
            \State Apply~\eqref{eq:reset_in}--\eqref{eq:reset_out}; clear Adam states for affected slices
          \EndFor
        \EndFor
      \EndIf
    \EndFor
  \EndFor
\EndFor
\end{algorithmic}
\end{algorithm}

\subsection{Theoretical Analysis}
\label{subsec:theory}

We analyze PRIME's tracking performance over the Dec-POMDP family $\{M_0, M_1, M_2\}$ formalized in Section~\ref{subsec:prelim}.
Because all $n$~agents share the same policy parameter $w_t \in \mathbb{R}^d$, the MAPPO surrogate objective at iteration~$t$ can be written as
\begin{equation}
\mathcal{L}_t(w) \;=\; \frac{1}{n}\sum_{u=1}^{n} \mathcal{L}_t^{(u)}(w),
\label{eq:shared_loss}
\end{equation}
where $\mathcal{L}_t^{(u)}$ is the clipped PPO loss evaluated on agent~$u$'s rollout data.
The gradient $\nabla\mathcal{L}_t(w) = \frac{1}{n}\sum_u \nabla\mathcal{L}_t^{(u)}(w)$ averages over the full multi-agent batch.

We adopt the following standard assumptions, which extend those used in the single-agent ReSiN analysis~\cite{he2025resin} to the shared-parameter CTDE setting.

\begin{assumption}[$L$-Smoothness]
\label{ass:smooth}
There exists $L > 0$ such that $\|\nabla\mathcal{L}_t(w) - \nabla\mathcal{L}_t(w')\| \leq L\|w - w'\|$ for all $w, w' \in \mathbb{R}^d$ and all $t$.
\end{assumption}

\begin{assumption}[$\mu$-Strong Convexity (Local)]
\label{ass:convex}
There exists $\mu > 0$ such that
$\langle \nabla\mathcal{L}_t(w) - \nabla\mathcal{L}_t(w'),\, w - w' \rangle
\;\geq\; \mu\,\|w - w'\|^2$
for all $w, w'$ within the 
PPO trust region at iteration~$t$.
This local condition is enforced by the clipped surrogate objective, 
which restricts each policy update to a neighborhood where the loss 
surface is approximately quadratic~\cite{schulman2017proximal,lascu2025ppo}.
\end{assumption}

\begin{assumption}[Bounded Non-Stationarity]
\label{ass:variation}
The cumulative drift of the phase-conditional optimum $w_t^* = \arg\min_w \mathcal{L}_t(w)$ across the eight transitions of the family $\{M_0, M_1, M_2\}$ is bounded: $P_T = \sum_{t=1}^{T-1}\|w_{t+1}^* - w_t^*\| < \infty$.
\end{assumption}

\begin{assumption}[Selective Reset Energy]
\label{ass:energy}
At each reset event, parameters in the silent set $\mathcal{S}_t = \bigcup_l \mathcal{S}_l$ of cardinality $d_t^* = |\mathcal{S}_t|$ are reinitialized.
The perturbation is modeled as a projection-restricted noise injection: $E_t = \eta\gamma_{\mathrm{r}}\Pi_t\zeta_t$ with $\zeta_t \sim \mathcal{N}(0,I_d)$, where $\Pi_t$ is the orthogonal projection onto the silent subspace of dimension $d_t^*$, and $\gamma_{\mathrm{r}} > 0$ controls the perturbation strength.
The expected perturbation energy satisfies $\mathbb{E}\|E_t\|^2 = \eta^2\gamma_{\mathrm{r}}^2 d_t^*$, with $d_t^* \leq d$~\cite{he2025resin}.
\end{assumption}

The PRIME update rule augments the standard PPO gradient step with the selective perturbation:
\begin{equation}
w_{t+1} \;=\; w_t \;-\; \eta\,\nabla\mathcal{L}_t(w_t) \;+\; \eta\gamma_{\mathrm{r}}\,\Pi_t\zeta_t.
\label{eq:update_rule}
\end{equation}
Between detection events the silent set is empty, so $\Pi_t = 0$ and~\eqref{eq:update_rule} reduces to the plain PPO step; the sums over $d_t^*$ below therefore accumulate only at the periodic reset events of Algorithm~\ref{alg:prime}.

\begin{theorem}[PRIME Dynamic Regret Bound]
\label{thm:prime}
Under Assumptions~\ref{ass:smooth}--\ref{ass:energy}, with learning rate $\eta \leq 1/L$, the time-averaged squared tracking error of the shared policy parameter satisfies
\begin{equation}
\frac{1}{T}\sum_{t=1}^{T}\mathbb{E}\|e_t\|^2 \;\leq\;
\frac{2}{\mu\eta T}\!\left(\mathbb{E}\|e_0\|^2 + \frac{2P_T^2}{\mu\eta}\right)
+ \frac{2\eta\gamma_{\mathrm{r}}^2}{\mu}\cdot\frac{1}{T}\sum_{t=1}^{T} d_t^*,
\label{eq:bound}
\end{equation}
where $e_t = w_t - w_t^*$ is the tracking error and $P_T = \sum_{t=1}^{T-1}\|w_{t+1}^* - w_t^*\|$ is the path length of the optimal parameters.
\end{theorem}

The bound takes the same form as ReSiN's single-agent tracking
analysis~\cite{he2025resin}: the multi-agent structure enters through the
team-averaged objective $\mathcal{L}_t$ and the path length $P_T$, and the
practical content lies in how small the intersection criterion keeps
$d_t^*$. The proof, together with the supporting
noise-energy lemma, is given in Appendix~B of the supplementary material.

\textbf{Comparison with blanket perturbation.}
Undiscriminated noise injection (e.g., shrink-and-perturb~\cite{nikishin2022primacy}) corresponds to $\Pi_t = I_d$ and $d_t^* = d$ for all $t$, yielding a perturbation term $\frac{2\eta\gamma_{\mathrm{r}}^2 d}{\mu}$.
Because the bidirectional intersection criterion~\eqref{eq:silent_set} ensures $d_t^* \ll d$ in practice, PRIME's perturbation term is smaller by a factor of $\bar{d}^*/d$, where $\bar{d}^* = T^{-1}\sum_t d_t^*$ is the time-averaged silent-subspace dimension.

\textbf{Role of phase non-stationarity.}
PRIME reduces the perturbation term in~\eqref{eq:bound} without affecting the drift term $P_T^2$, improving tracking by lowering the cost of plasticity maintenance.
The experiments in Section~\ref{sec:experiments} confirm this: PRIME's dormant fraction stays at $10$--$20\%$ while MAPPO reaches $40$--$45\%$, so $\bar{d}^*$ is a small fraction of $d$.
Appendix~C of the supplementary material discusses the assumptions underlying Theorem~\ref{thm:prime} in more detail, including how the additive-noise reset model relates to PRIME's actual reset operations.

\section{Experiments}
\label{sec:experiments}

\subsection{Setup}
\textbf{Environment.}
We evaluate on a $1\,\text{km} \times 1\,\text{km}$ post-disaster urban grid with $U{=}3$~UAVs serving $N{=}20$~ground users organized in $G{=}3$~RPGM groups, with slot duration $\Delta t{=}60\,$s and an episode horizon of $32$~steps.
Fixed simulation parameters and the rotary-wing energy model are listed in the upper part of Table~\ref{tab:env_params}.
Two training regimes follow Section~\ref{sec:motivation}: \textit{Normal} mode holds a single fixed phase for $1{,}500$~iterations (${\sim}3 \!\times\! 10^6$ environment steps), and \textit{Change} mode cycles $\{0,1,2\}$ three times ($9{,}000$~iterations, ${\sim}18.4 \!\times\! 10^6$ steps, $1{,}000$~iterations per phase, eight transitions).
Each phase simultaneously alters user mobility speed, service radius, and the demand pattern with its overlap penalty $\lambda_{\mathrm{ov}}$ (lower part of Table~\ref{tab:env_params}); these concurrent shifts create the non-stationarity that motivates PRIME's plasticity maintenance.

\begin{table}[!t]
\renewcommand{\arraystretch}{1.3}
\caption{UAV-ECN Simulation Parameters}
\label{tab:env_params}
\centering
\begin{tabular}{l l}
\hline
\textbf{Parameter} & \textbf{Value} \\
\hline
\multicolumn{2}{l}{\emph{Fixed across all phases}} \\
\hline
Area size $S \times S$ & $1000\,\text{m} \times 1000\,\text{m}$ \\
Number of UAVs $U$ & 3 \\
Number of ground users $N$ & 20 \\
RPGM groups $G$ & 3 \\
Slot duration $\Delta t$ & 60\,s \\
Episode horizon & 32 steps \\
Max horizontal speed & 12.0\,m/s \\
Action step distance & 100.0\,m \\
Min inter-UAV distance $d_{\min}$ & 10.0\,m \\
Max users per UAV & 5 \\
Frequency reuse factor & 1 \\
Adjacent-channel leakage $\rho_{\mathrm{adj}}$ & $10^{-3}$ \\
Gauss--Markov $\alpha$ / $\sigma$ & 0.9 / 0.08 \\
RPGM follow gain $\kappa$ & 0.3 \\
Rotary-wing $P_0$ / $P_i$ & 90\,W / 110\,W \\
Tip speed $U_{\mathrm{tip}}$ & 120\,m/s \\
\hline
\multicolumn{2}{l}{\emph{Phase-specific (Change mode)}} \\
\hline
 & Phase 0 \;/\; Phase 1 \;/\; Phase 2 \\
\hline
User speed (normalized) & 0.2 \;/\; 0.5 \;/\; 0.8 \\
Service radius (m) & 200 \;/\; 150 \;/\; 150 \\
Demand pattern & all L \;/\; half$\to$M \;/\; half$\to$H \\
Demand thresholds (Mbps) & L: 0.5,\; M: 1.0,\; H: 2.0 \\
Reward weights & L: 5,\; M: 10,\; H: 20 \\
Overlap penalty $\lambda_{\mathrm{ov}}$ & 20 \;/\; 40 \;/\; 80 \\
\hline
\multicolumn{2}{l}{\emph{Training schedule}} \\
\hline
Phase sequence (Change mode) & $\{0,1,2\} \times 3$ repeats \\
Iterations per phase (Change) & 1{,}000 \\
Total iterations (Normal / Change) & 1{,}500 / 9{,}000 \\
\hline
\end{tabular}
\end{table}

\textbf{Baselines.}
We compare three methods sharing the same shared two-layer MLP architecture ($H{=}32$, ReLU) and MAPPO training pipeline, with the plasticity strategy as the only difference:
(i)~\textbf{MAPPO}, no plasticity mechanism;
(ii)~\textbf{PRIME-Forward}, which resets all neurons with $\hat{s}_{l,i} \leq \tau_d$ regardless of gradient magnitude, subsuming the forward-only family (ReDo~\cite{sokar2023dormant}, shrink-and-perturb~\cite{nikishin2022primacy}, Plasticity Injection~\cite{nikishin2024deep});
and (iii)~\textbf{PRIME}, the bidirectional intersection criterion with $\tau_d{=}0.5$, $\tau_g{=}0.08$, and reset period $F{=}200$ mini-batch steps.
PRIME constitutes the multi-agent CTDE adaptation of ReSiN's bidirectional criterion~\cite{he2025resin} with the shared-batch aggregation in Section~\ref{subsec:detection}.
The full hyperparameter list is provided in Appendix~A of the supplementary material.

\textbf{Evaluation protocol.}
We report episode return, coverage rate (fraction of users receiving above-threshold data rate), served-user count, collision rate, and energy usage rate, using interquartile mean (IQM)~\cite{agarwal2021deep} as the primary scalar metric for robustness to outlier episodes.
The evaluation focuses on a single ECN scenario ($U{=}3$, $H{=}32$) so that the effect of bidirectional detection is isolated under a controlled non-stationarity profile with clearly identifiable transition points.
The small network ($1{,}184$ trainable actor parameters) places a high premium on detection precision, since every unnecessary reset removes a significant share of capacity.
PRIME's mechanism is architecture- and task-agnostic, requiring no environment-specific tuning beyond $\tau_d$ and $\tau_g$ (robustness confirmed by ablation, Fig.~\ref{fig:ablation_threshold}); extensions to larger teams and networks, standard MARL benchmarks (SMAC~\cite{samvelyan2019starcraft}, MPE~\cite{lowe2017multi}), and bootstrap confidence intervals over larger seed populations~\cite{agarwal2021deep} are left to future work.
All experiments are repeated with two random seeds ($42$ and $43$) under deterministic PyTorch settings.
Reported IQM values are the mean of per-seed IQMs; curves with shaded regions show the cross-seed mean and the min--max band across seeds, ablation points carry min--max error bars, and the remaining per-neuron diagnostic traces are from a single representative seed ($42$).

\subsection{Task and ECN Performance}

\textbf{Stationary baseline.}
In the stationary Normal setting, vanilla MAPPO attains the highest IQM return ($83.392$), followed by PRIME ($78.352$) and PRIME-Forward ($50.622$).
PRIME's periodic resets trade a small amount of steady-state performance for sustained representational health (DNF at $10$--$18\%$ versus MAPPO's $40$--$45\%$), an investment whose value becomes clear under non-stationarity.
Full Normal-mode curves and bar charts are reported in Appendix~E of the supplementary material.

\begin{figure}[!t]
\centering
\subfloat[Training reward curves]{
  \includegraphics[width=0.43\columnwidth]{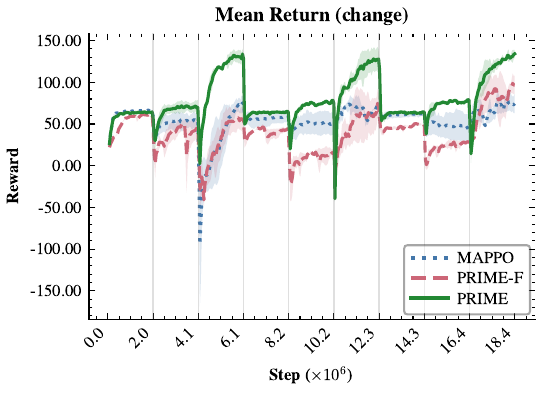}
  \label{fig:reward_change}}
\hfill
\subfloat[Normalized IQM across five ECN metrics]{
  \includegraphics[width=0.47\columnwidth]{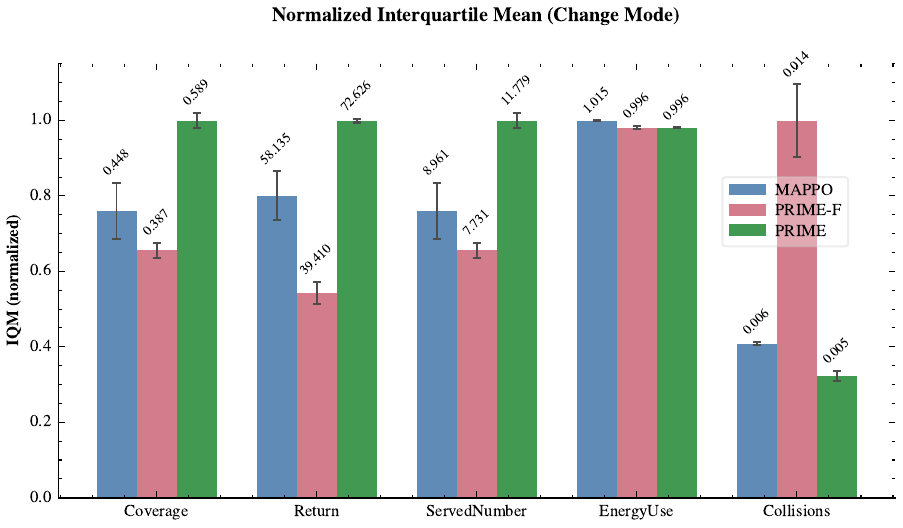}
  \label{fig:iqm_change}}
\caption{Change-mode task performance.
(a)~Across eight phase transitions (gray lines), PRIME recovers to the highest within-phase reward; PRIME-Forward suffers progressively deeper post-switch dips.
(b)~PRIME leads on return, coverage, and served users, with the lowest collision rate among the compared methods.}
\label{fig:task_change}
\end{figure}

\textbf{Change mode.}
Under cyclic phase switching, the ordering reverses (Fig.~\ref{fig:reward_change}).
All three methods dip at each phase boundary, but they differ in dip depth and recovery speed.
PRIME recovers to within-phase peaks of ${\approx}\,140$ in each phase, whereas MAPPO peaks near $85$ and PRIME-Forward near $80$ in later phases.
However, PRIME-Forward suffers much deeper post-switch dips than MAPPO, and the dips deepen progressively, reflecting compounding disruption from indiscriminate resets.

In terms of IQM return (Fig.~\ref{fig:iqm_change}), PRIME reaches $72.626$, compared with $58.135$ for MAPPO ($+24.9\%$) and $39.410$ for PRIME-Forward.
The PRIME-Forward gap is the more informative one: forward-only resets destroy gradient-active neurons precisely when the optimizer needs stable learning signals to adapt to a new phase.

\begin{figure*}[!t]
\centering
\subfloat[Coverage rate]{
  \includegraphics[width=0.28\textwidth]{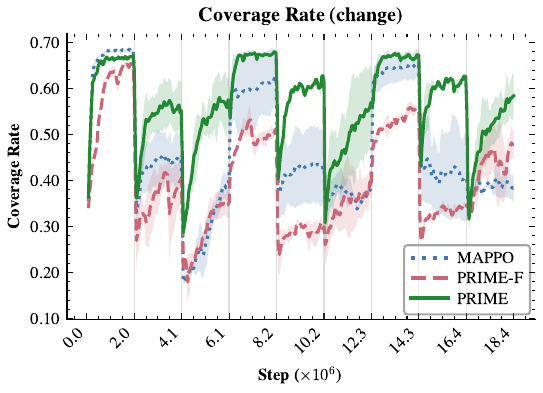}
  \label{fig:coverage_change}}
\hfill
\subfloat[Served users]{
  \includegraphics[width=0.28\textwidth]{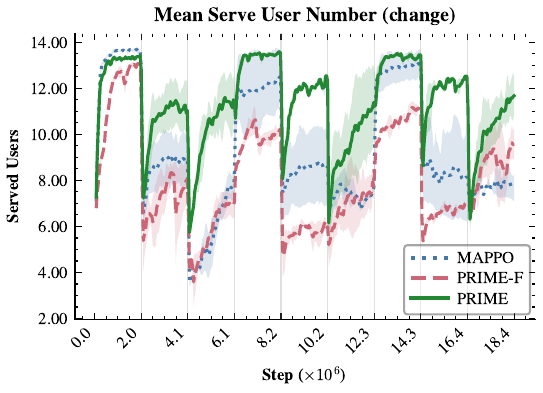}
  \label{fig:served_change}}
\hfill
\subfloat[Collisions]{
  \includegraphics[width=0.28\textwidth]{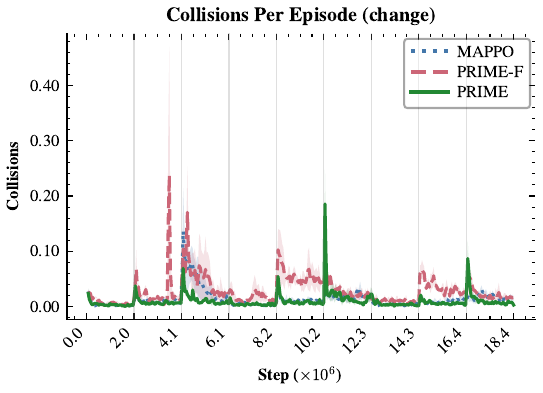}
  \label{fig:collisions_change}}
\caption{ECN performance in Change mode: (a)~coverage rate, (b)~served users, and (c)~collision rate; PRIME is lowest among the compared methods on (c).}
\label{fig:ecn_metrics}
\end{figure*}

\textbf{ECN metrics.}
Figure~\ref{fig:ecn_metrics} reports three application-level metrics.
PRIME recovers coverage fastest after each transition (${\sim}\,0.68$ at within-phase peaks, versus ${\sim}\,0.60$ for MAPPO and ${\sim}\,0.45$ for PRIME-Forward in later phases) and serves the most users (${\sim}\,14$ per slot at peak, versus ${\sim}\,9$--$11$); PRIME-Forward's deepest post-switch dips (${\sim}\,4$--$5$ served users) reflect the transient disruption of over-resetting.
All methods produce brief collision spikes at phase boundaries as trajectories are re-planned; among the compared methods, PRIME maintains the lowest collision rate (IQM $= 0.005$), suggesting that the selective reset criterion preserves neurons responsible for inter-UAV separation while reclaiming genuinely inactive capacity.

\subsection{Plasticity Diagnostics}

We examine four neural-network-level indicators that explain the performance differences above.

\begin{figure*}[!t]
\centering
\subfloat[Total dormant fraction]{
  \includegraphics[width=0.28\textwidth]{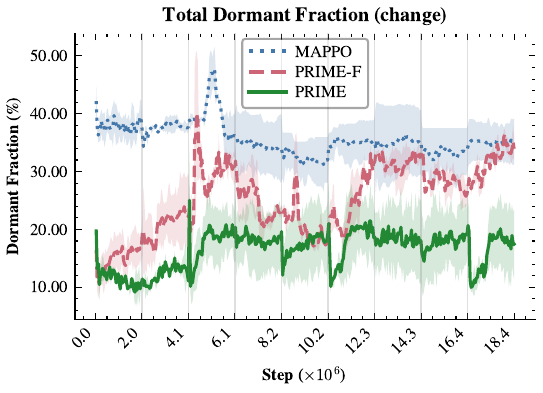}
  \label{fig:dormant_total}}
\hfill
\subfloat[Average feature rank]{
  \includegraphics[width=0.28\textwidth]{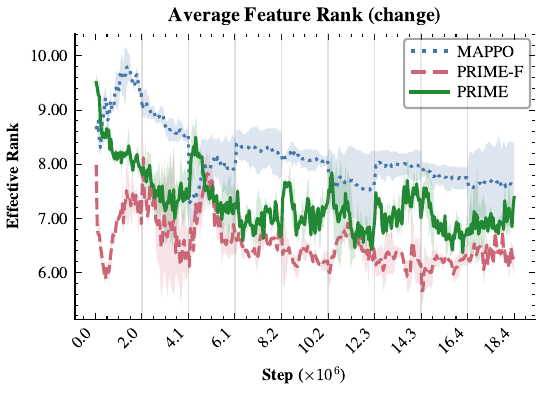}
  \label{fig:rank}}
\hfill
\subfloat[Policy entropy]{
  \includegraphics[width=0.28\textwidth]{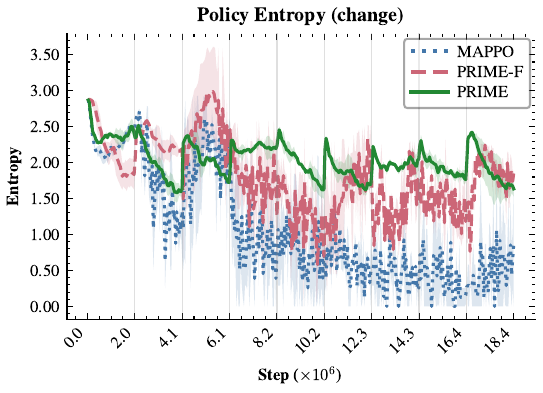}
  \label{fig:entropy}}
\caption{Plasticity diagnostics (Change mode).
(a)~PRIME maintains the lowest dormant fraction (${\sim}\,10$--$20\%$) through periodic selective resets.
(b)~MAPPO's high rank is inflated by noisy dormant-neuron activations; PRIME's lower rank reflects a more coherent representation.
(c)~PRIME retains higher entropy, preserving exploration across phase transitions.}
\label{fig:diagnostics}
\end{figure*}

\textbf{Dormant neuron fraction.}
Figure~\ref{fig:diagnostics}(a) shows that PRIME maintains the lowest DNF throughout Change-mode training (${\sim}\,10$--$20\%$), compared with MAPPO (${\sim}\,40$--$45\%$) and PRIME-Forward (${\sim}\,15$--$30\%$).
PRIME's DNF exhibits phase-synchronous oscillations: a brief rise at each phase boundary followed by a drop upon the next reset sweep.
This pattern confirms that the bidirectional criterion targets genuinely silent neurons at each phase transition.

\textbf{Feature rank.}
MAPPO retains the highest effective rank (${\approx}\,8$--$10$; Fig.~\ref{fig:diagnostics}(b)).
This may appear favorable, but is consistent with noisy activations from its ${\sim}\,40\%$ dormant neurons inflating statistical diversity in the feature matrix without a corresponding gain in policy quality.
PRIME's rank (${\approx}\,7$--$8$) is lower but corresponds to a more coherent representation where every active neuron contributes to the output.
PRIME-Forward's rank collapses to ${\approx}\,5$--$6$ over time, confirming that indiscriminate resets destroy learned structure faster than the optimizer can rebuild it.

\textbf{Policy entropy.}
PRIME sustains higher entropy across phase transitions (Fig.~\ref{fig:diagnostics}(c)), retaining the exploratory capacity needed to adapt when the objective changes.
MAPPO's entropy declines monotonically, consistent with a shrinking effective action space as dormancy accumulates.

\textbf{Silent neuron decomposition.}
Figure~S10 in the supplementary material decomposes PRIME's neuron population into three sets per layer: forward-dormant ($\mathcal{D}_l$), backward-silent ($\mathcal{G}_l$), and their intersection ($\mathcal{S}_l = \mathcal{D}_l \cap \mathcal{G}_l$).
Policy Layer~0 is the highest-dormancy layer under PRIME, consistent with Section~\ref{subsec:plasticity_loss}'s finding that input-adjacent layers accumulate dormancy most aggressively; the team-aggregate DNF reported as $10$--$20\%$ in Fig.~\ref{fig:diagnostics}(a) averages across all four hidden layers and is pulled down by the lower-dormancy value layers.
In Policy Layer~0 specifically, $|\mathcal{G}_l|/H_l \approx 60$--$80\%$, $|\mathcal{D}_l|/H_l \approx 30$--$40\%$, and $|\mathcal{S}_l|/H_l \approx 20$--$30\%$.
Two facts follow.
First, $|\mathcal{D}_l \setminus \mathcal{G}_l|/|\mathcal{D}_l| \approx 20$--$30\%$ of forward-dormant neurons are preserved by the bidirectional criterion because they still receive non-trivial gradient signal.
Second, this preserved fraction is the empirical counterpart to the forward-only false-positive lower bound of $50$--$90\%$ measured on vanilla MAPPO in Section~\ref{subsec:decoupling}: PRIME's bidirectional gate brings the equivalent rate of would-be over-resets down by a factor of two to three, directly confirming the diagnostic from Section~\ref{sec:motivation}.

\subsection{Ablation Studies}

We ablate seven design choices in Change mode: the two detection thresholds and the reset period (with PRIME-Forward as the forward-only reference), phase-signal dependence, and the three design-axis substitutions of Table~\ref{tab:ablation_signal} (backward-signal source, aggregation granularity, and intervention operator).

\begin{figure}[!t]
\centering
\includegraphics[width=0.8\columnwidth]{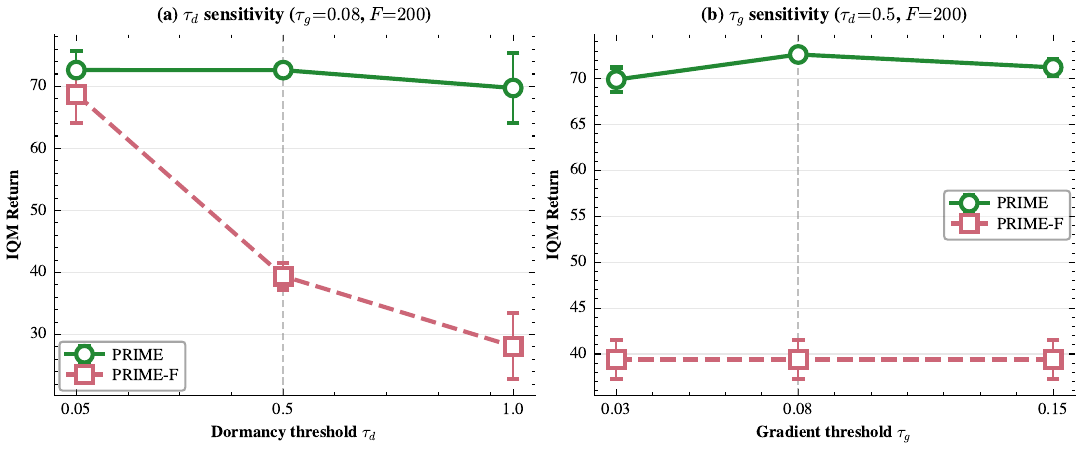}
\caption{Threshold sensitivity (Change mode).
(a)~IQM return vs.\ $\tau_d \in \{0.05, 0.5, 1.0\}$ ($\tau_g{=}0.08$, $F{=}200$).
PRIME is robust across the range; PRIME-Forward degrades sharply at larger $\tau_d$.
(b)~IQM return vs.\ $\tau_g \in \{0.03, 0.08, 0.15\}$ ($\tau_d{=}0.5$, $F{=}200$).
PRIME improves with moderately permissive $\tau_g$; PRIME-Forward is invariant.}
\label{fig:ablation_threshold}
\end{figure}

\textbf{Dormancy threshold $\tau_d$.}
We vary $\tau_d \in \{0.05, 0.5, 1.0\}$ with $\tau_g{=}0.08$ and $F{=}200$ fixed (Fig.~\ref{fig:ablation_threshold}(a)).
PRIME's IQM return stays above $64$ across the full range, with only a mild decrease at $\tau_d{=}1.0$ where the dormancy criterion becomes overly inclusive.
PRIME-Forward drops from ${\sim}\,69$ at $\tau_d{=}0.05$ (few neurons qualify) to ${\sim}\,28$ at $\tau_d{=}1.0$ (nearly all neurons are reset).
This divergence illustrates the protective effect of the gradient-silence criterion: even when the forward threshold admits many neurons, the backward filter prevents PRIME from resetting those that are still receiving useful gradient signal.

\textbf{Gradient threshold $\tau_g$.}
Varying $\tau_g \in \{0.03, 0.08, 0.15\}$ with $\tau_d{=}0.5$ fixed (Fig.~\ref{fig:ablation_threshold}(b)), PRIME's IQM return rises from ${\sim}\,70$ at $\tau_g{=}0.03$ to $72.626$ at the default $\tau_g{=}0.08$, and dips slightly to ${\sim}\,71$ at $0.15$.
A moderately permissive gradient threshold captures the most beneficial resets without over-inclusion.
PRIME-Forward is invariant to $\tau_g$ (IQM $= 39.410$ at all three values), since it ignores gradient information by construction.

\begin{figure}[!t]
\centering
\subfloat[Reset period $F$]{
  \includegraphics[width=0.43\columnwidth]{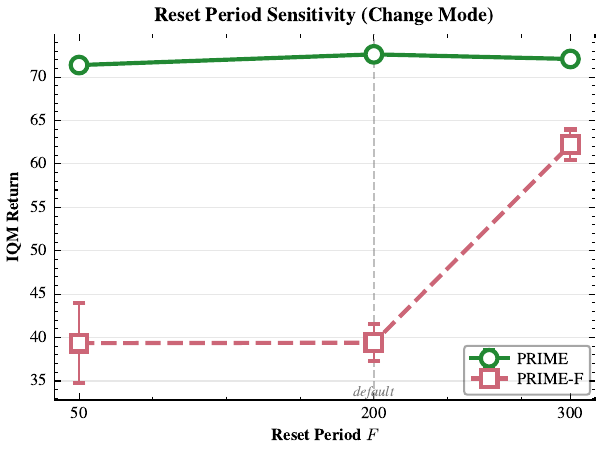}
  \label{fig:ablation_F}}
\hfill
\subfloat[Phase-signal independence]{
  \includegraphics[width=0.47\columnwidth]{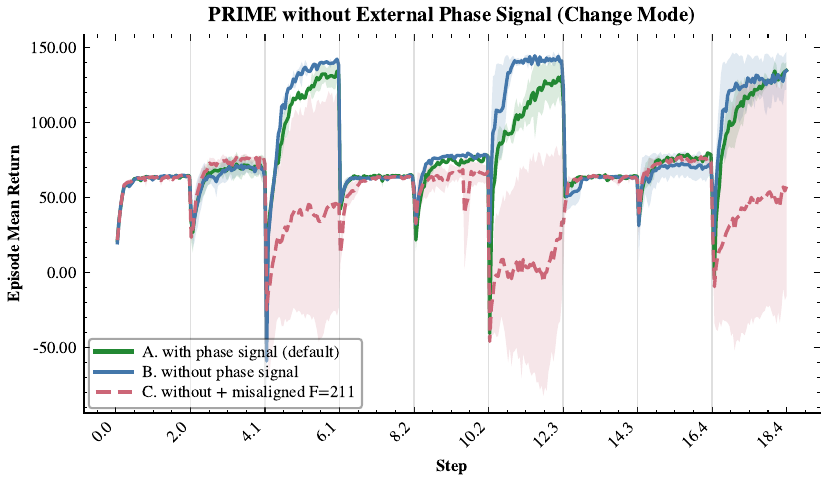}
  \label{fig:ablation_alignment}}
\caption{Reset-schedule ablations (Change mode).
(a)~IQM return for $F \in \{50, 200, 300\}$: PRIME is stable across reset periods; PRIME-Forward improves with less frequent resets.
(b)~Three reset-schedule configurations: A: default schedule from the main experiments, which augments periodic detection ($F{=}200$) with a full reset sweep at each phase boundary; B: pure periodic detection at $F{=}200$, as specified by Algorithm~\ref{alg:prime}; C: pure periodic detection at $F{=}211$.
PRIME's IQM return is comparable in A and B and degrades gracefully in C, confirming that the periodic detection mechanism alone is sufficient.}
\label{fig:ablation_period_alignment}
\end{figure}

\textbf{Reset period $F$.}
With $\tau_d{=}0.5$ and $\tau_g{=}0.08$, we test $F \in \{50, 200, 300\}$ (Fig.~\ref{fig:ablation_period_alignment}(a)).
PRIME's IQM return stays within $71$--$73$ across all three values, confirming that the intersection criterion is selective enough to prevent harmful resets even at high frequency.
PRIME-Forward is sensitive to $F$: its IQM rises from ${\sim}\,39$ at $F{=}50$ (frequent over-resetting) to ${\sim}\,62$ at $F{=}300$ (less frequent resets allow partial recovery between interventions).

\textbf{Phase-signal independence.}
Algorithm~\ref{alg:prime} specifies a purely periodic mechanism ($c \bmod F = 0$, no knowledge of phase boundaries), but the main experimental runs (variant A in Fig.~\ref{fig:ablation_period_alignment}(b)) additionally trigger a full reset sweep at each boundary; whether the periodic-only mechanism suffices is therefore an empirical question, answered by comparing three configurations.
Variant B removes the phase-triggered sweep, leaving exactly the procedure of Algorithm~\ref{alg:prime}; variant C further shifts the period to a co-prime $F{=}211$, so the offset between resets and regime changes drifts across phases and is effectively randomized.
PRIME's IQM return is $72.626$, $75.301$, and $55.790$ under A, B, and C respectively; coverage rates follow the same ordering ($0.589$, $0.617$, $0.520$).
Variant B matches and slightly exceeds the default, demonstrating that the phase-triggered augmentation is not required---the periodic detection in Algorithm~\ref{alg:prime} suffices on its own.
Variant C degrades gracefully rather than failing: with the schedule intentionally adversarial to the phase structure, performance stays comparable to MAPPO ($58.135$) and well above the forward-only baseline, though with the largest cross-seed spread.
Together these results verify the design claim in Section~\ref{sec:method}: PRIME's plasticity recovery is decoupled from any external signal about when the environment changes.

The remaining three ablations vary choices that Algorithm~\ref{alg:prime}
fixes implicitly: where the backward signal originates, over whose experience
the detection statistics aggregate, and which operator acts on the silent
set. Each variant changes exactly one choice while keeping thresholds,
layer-wise normalization, the intersection gate, and the period $F{=}200$
identical to the main configuration. Table~\ref{tab:ablation_signal} reports
Change-mode results; the Seeds column distinguishes two-seed protocol values
from single-seed sweep points.

\begin{table}[!t]
\renewcommand{\arraystretch}{1.3}
\setlength{\tabcolsep}{3pt}
\caption{Design-Axis Ablations of the Detection Pipeline (Change Mode)}
\label{tab:ablation_signal}
\centering
\footnotesize
\begin{tabular}{l l c c l}
\hline
Variant & Modified choice & $\tau_g$ & Seeds & Return IQM \\
\hline
MAPPO & no reset & -- & 42,\,43 & 58.135 \\
PRIME-Forward & \begin{tabular}[t]{@{}l@{}}gradient gate\\ removed\end{tabular} & -- & 42,\,43 & 39.410 \\
PRIME & none (reference) & 0.08 & 42,\,43 & 72.626 {\scriptsize[72.2,\,73.0]} \\
\hline
AuxGrad & \begin{tabular}[t]{@{}l@{}}signal $\to$\\ output gradient\end{tabular} & 0.03 & 42 & 34.095 \\
 & & 0.08 & 42 & 45.146 \\
 & & 0.15 & 42,\,43 & 73.439 {\scriptsize[65.7,\,81.2]} \\
 & & 0.30 & 42 & 71.079 \\
SingleSlice & \begin{tabular}[t]{@{}l@{}}statistics $\to$\\ agent-0 slice\end{tabular} & 0.08 & 42,\,43 & 64.480 {\scriptsize[60.0,\,69.0]} \\
Noise & \begin{tabular}[t]{@{}l@{}}reset $\to$\\ perturbation\end{tabular} & 0.08 & 42,\,43 & 74.414 {\scriptsize[73.0,\,75.8]} \\
\hline
\end{tabular}
\end{table}

\textbf{Backward-signal source.}
PRIME's gradient gate reads the gradient that the PPO training loss has
already deposited on the weights. The \emph{AuxGrad} variant replaces this
signal with the loss-independent output-sensitivity gradient used by
ReSiN~\cite{he2025resin}, $g_{W} = \nabla_{W} \sum_{n} f_{W}(x_{n})$,
obtained from one additional forward--backward pass per network at each
detection event; the per-neuron reduction, layer-wise normalization,
intersection gate, and reset operator are unchanged. Because this proxy is
not scaled by the optimizer that produced it, its threshold requires
calibration, and we sweep $\tau_g^{\mathrm{aux}} \in \{0.03, 0.08, 0.15,
0.30\}$ on seed 42 before committing to a two-seed comparison at the best
value. The sweep spans 47 IQM points: 34.095, 45.146, 81.208, and 71.079
respectively, with two of the four settings falling below vanilla MAPPO
(58.135). Over the same grid, the training-gradient signal varies by roughly
2.6 points (Fig.~\ref{fig:ablation_threshold}(b)). Cumulative reset counts
track the swing. At $0.03$ the gate essentially never opens: 5{,}012 resets
over the full run, 3.9\% of PRIME's volume on the same seed (127{,}492), so
the variant degenerates to
vanilla MAPPO with detection overhead. At $0.08$ the reset schedule
repeatedly stalls, flattening after roughly 10M steps at 49.8\% of PRIME's
volume, while the value-network zero-activation fraction spikes above 60\%
inside the stalled windows. At $0.15$ clearing is continuous ($1.27\times$
PRIME's volume) and performance reaches parity. The tuned threshold does not
transfer across seeds, however. On seed 43 the same setting reproduces the
stalling mode, with no interventions after roughly 11M steps and 48\% of the
seed-42 reset volume, and lands at 65.669, which is 6.6 points below PRIME's
weaker seed. The resulting two-seed mean, 73.439, carries a 15.5-point seed
range, against 0.8 points for PRIME (72.246 and 73.006). Cumulative volume
does not separate the two signals: PRIME's own count varies from 127{,}492
to 231{,}085 across its seeds while performance moves by less than a point.
What separates them is whether the gate stays responsive. PRIME clears
continuously on both seeds; the proxy's gate closes outright on one of them,
and performance falls with it. In the stationary Normal mode the
Change-tuned threshold transfers: AuxGrad reaches 77.488 against PRIME's
78.352, with a 0.04-point seed band; we note that the Change-mode stalls
emerged only after roughly 10M steps, beyond the 3.07M-step horizon of the
stationary runs. The training gradient thus costs no additional computation,
arrives pre-scaled by the optimizer, and holds performance within 2.6 points
over the threshold grid and 0.8 points across seeds; the output-sensitivity
proxy reaches parity only after a per-environment threshold search and
remains seed-fragile at its best setting under non-stationarity.

\textbf{Aggregation granularity.}
Section~\ref{subsec:cross_agent} measured $5$--$18\%$ cross-agent
disagreement in per-agent dormancy statistics; the \emph{SingleSlice} variant
quantifies the behavioral cost of acting on one agent's view. Both criteria
are computed from agent~0's $(B{\times}1)$ slice: forward statistics from
that agent's rows, and the gradient from the PPO loss restricted to the same
rows, while resets still apply to the shared network. This emulates a
single-agent detector operating inside a shared-parameter system. The variant
keeps part of the benefit, 64.480 against MAPPO's 58.135, but gives up 8.1
IQM points relative to PRIME, an 11.2\% drop, with disjoint per-seed ranges:
its stronger seed (68.957) remains below PRIME's weaker one (72.246).
Coverage falls from 0.589 to 0.543 while collisions hold at PRIME's 0.005, so
the cost concentrates in service quality rather than safety. The disagreement
measured in Section~\ref{subsec:cross_agent} is the phenomenon; the 8.1-point
gap is its consequence: team-level aggregation is not only the well-posed
choice but a measurably better one.

\textbf{Intervention operator.}
The \emph{Noise} variant keeps detection unchanged and replaces the
output-preserving reset with the perturbation family used by
ReSiN~\cite{he2025resin}: additive Gaussian noise with standard deviation
$0.1\times$ the layer's initialization bound, applied to the incoming weights
and bias of the silent set with no outgoing zeroing, at the matched cadence
$F{=}200$. Performance matches PRIME: 74.414 against 72.626, with overlapping
per-seed ranges and near-identical intervention volumes across seeds
(148{,}850 and 148{,}041, within 0.6\%, versus 127{,}492 for PRIME on the
same seed). The roughly $1.17\times$ higher volume is consistent with
perturbation lifting a neuron out of dormancy more gradually than
reinitialization, so that some neurons are treated repeatedly. Two
conclusions follow. The performance load rests on detection: once the
intersection criterion selects the neurons, hard reset and stochastic
perturbation are interchangeable at matched cadence. And the main-experiment
gains do not depend on the output-preserving convention; we retain the hard
reset as the default for its output continuity at the moment of
intervention.

\textbf{Activation function.}
All experiments use ReLU, whose exact zeros are the natural setting for dormancy detection; for smooth rectifiers (Leaky-ReLU, GELU, Swish) the normalized threshold $\tau_d$ already implements near-zero-magnitude detection, and the gradient criterion is activation-agnostic by construction.
We therefore expect PRIME to transfer, with the optimal $\tau_d$ possibly shifting; empirical validation is left to future work.

\textbf{Summary.}
Across all seven ablation axes (dormancy threshold, gradient threshold,
reset period, phase-signal independence, backward-signal source, aggregation
granularity, and intervention operator), PRIME is robust to every
implementation choice, while the sensitivities that remain are exactly the
claimed ones. Substituting the training-gradient signal exposes a 47-point
threshold sensitivity and a 15.5-point seed spread; restricting aggregation
to a single agent's slice costs 8.1 IQM points; exchanging the reset operator
for a matched-cadence perturbation costs nothing; and wherever the
forward-only criterion admits many resets, PRIME-Forward degrades sharply
while PRIME does not.
This confirms the central claim of Section~\ref{subsec:theory}: the intersection criterion $\mathcal{S}_l = \mathcal{D}_l \cap \mathcal{G}_l$ restricts resets to the genuinely expendable subspace, keeping the perturbation term $\bar{d}^*$ in the regret bound~\eqref{eq:bound} small.
Neurons in $\mathcal{D}_l \setminus \mathcal{G}_l$ carry non-zero gradients, are actively participating in learning, and must not be reset.

\subsection{Computational Overhead}

PRIME adds three costs to vanilla MAPPO: per-neuron activation logging during the forward pass, an $\ell_1$ reduction over each neuron's weight gradients at detection events (the gradients themselves are reused from the PPO backward pass), and the reset writes, which touch $|\mathcal{S}_l| \approx 2$--$5$ of $H{=}32$ neurons per layer in practice (Fig.~S13 in the supplementary material).
In our setting ($H{=}32$, $n{=}3$, $F{=}200$), wall-clock training time increases by approximately $3.8\%$ over MAPPO ($23$h$12$m vs.\ $22$h$21$m for $18.4{\times}10^6$ environment steps on the same hardware).
Per-neuron scoring is $O(H \cdot d_{l,\mathrm{in}})$ per layer, dominated by the PPO forward and backward passes that scale identically, and the detection logic operates on per-neuron aggregates, adding no cost in $n$ beyond the batch aggregation itself.
Memory overhead is two floats per neuron ($512$ bytes for the current architecture).

\section{Related Work}
\label{sec:related}

\subsection{UAV-Assisted ECN and Non-Stationarity}

UAV swarms for post-disaster connectivity restoration have been studied from trajectory optimization~\cite{wu2018joint,zeng2019energy} to energy-aware scheduling and user association~\cite{sun2023energy,hussain2024computing}; Zhao et al.~\cite{zhao2025survey} provide a recent survey of deep reinforcement learning (DRL)-based UAV communications.
Single-agent DRL has been applied to adaptive placement and spectrum management~\cite{badnava2021spectrum}, and cooperative MARL formulations extend this to multi-UAV coordination using CTDE architectures~\cite{foerster2018counterfactual,yuan2025multiagent}.
PE-MAMoE~\cite{qiu2026plasticity} introduces sparse MoE routing with phase-aware noise injection for objective switching.
These works, however, assume that the policy network retains its representational capacity across phase boundaries.

When the environment does change, MARL methods typically handle non-stationarity by modeling or anticipating the external shift.
Experience-replay stabilization~\cite{foerster2017stabilising} and opponent modeling~\cite{foerster2018learning,zhao2022pola} address the policy-drift form of non-stationarity that arises from co-learning agents.
Meta-RL~\cite{finn2017maml,al2018continuous} targets task-distribution shift but requires representative meta-training tasks.
Wang et al.~\cite{wang2025macph} recently proposed MACPH for non-stationary environments with unknown change points, combining composite replay buffers with adaptive parameter-space noise; the noise, however, is applied globally to actor parameters without neuron-level selectivity.
Galashov et al.~\cite{galashov2024nonstationary} introduced Soft Reset, which models parameter drift explicitly and pulls weights toward their initialization, but the reset is uniform across all parameters rather than targeted at silent neurons.
A common limitation runs through these approaches: they assume, implicitly or explicitly, that environmental change can be tracked through external signals such as reward shifts, transition changes, or task labels.
None addresses the network-level capacity degradation that compounds across phase switches, the problem that PRIME targets.

\subsection{Neural Plasticity Loss and Reset Methods}

Plasticity loss has been studied along two axes: diagnosis and mitigation.
On the diagnostic side, Dohare et al.~\cite{dohare2024loss} established that continual training degrades deep networks to the point where they offer no advantage over single-layer models.
Sokar et al.~\cite{sokar2023dormant} formalized the dormant neuron fraction as a proxy for capacity loss and proposed the ReDo algorithm, which reinitializes neurons whose forward activations fall below a threshold.
Lyle et al.~\cite{lyle2023understanding,lyle2024disentangling} connected this neuron-level phenomenon to a geometric signature at the representation level: the effective dimensionality of learned features shrinks as dormancy accumulates.
Juliani and Ash~\cite{juliani2024plasticity} showed that on-policy PPO is equally susceptible, with several off-policy fixes such as CReLU and plasticity injection~\cite{nikishin2024deep} failing to transfer.

On the mitigation side, existing methods span a spectrum from coarse to fine-grained.
Global perturbation approaches, including continual backpropagation~\cite{dohare2024loss}, the primacy-bias reset~\cite{nikishin2022primacy}, weight clipping~\cite{elsayed2024weight}, and regenerative regularization~\cite{kumar2023maintaining}, perturb or constrain the full parameter space without distinguishing active neurons from inactive ones.
MoE architectures~\cite{shazeer2017outrageously} can reduce dormancy through structural diversity~\cite{willi2024mixture}, but add routing overhead and load-balancing losses that complicate on-policy training~\cite{zhou2022mixture}.
ReDo~\cite{sokar2023dormant} and plasticity injection~\cite{nikishin2024deep} are more targeted, resetting or augmenting individual neurons, yet both rely solely on forward-activation scores and cannot distinguish genuinely expendable neurons from those receiving strong gradient signal during phase adaptation.
ReSiN~\cite{he2025resin} introduced the Silent Neuron activity index, combining forward dormancy with backward gradient silence into a single criterion: only neurons inactive in both directions qualify for reset.
The method was validated on single-agent adaptive video streaming with tracking-error guarantees.
ReSiN itself builds on PA-MoE~\cite{he2025pamoe}, which addressed plasticity under QoE shifts in the same setting through a plasticity-aware mixture-of-experts controller.

In the multi-agent setting, the dormancy problem takes a distinct form.
Qin et al.~\cite{qin2024dormant} found that dormant neurons in QMIX concentrate in the mixing network and that their prevalence grows with team size; their ReBorn method redistributes weights from over-active to dormant neurons through a monotonicity-aware transfer.
Yuan et al.~\cite{yuan2025multiagent} showed that coordination skills degrade across regime shifts even when individual performance is maintained, and proposed progressive task contextualization to mitigate inter-agent forgetting.
Neither ReBorn nor progressive contextualization uses backward gradient information to guide resets.
PRIME extends the bidirectional criterion of ReSiN to shared-parameter CTDE.
The extension is not cosmetic: Section~\ref{subsec:cross_agent} shows that per-agent dormancy disagreement reaches $5$--$18\%$ in policy layers under vanilla MAPPO, so the single-rollout activation and gradient statistics used by ReSiN do not directly transfer to multi-agent training without team-level aggregation.
PRIME instead computes activation and gradient scores over the full $B{\times}n$ team-aggregated batch, which both eliminates per-agent inconsistency and exploits the richer statistics afforded by cooperative training.
Section~\ref{sec:experiments} makes the comparison empirical: ReSiN's
output-sensitivity signal reaches PRIME's Change-mode performance only after
a per-environment threshold sweep and loses it to gate closure on one seed,
while ReSiN's perturbation operator, paired with PRIME's detector at matched
cadence, performs on par with the hard reset.
To our knowledge, PRIME is the first method in multi-agent communication systems to detect non-stationarity through internal neuron-level plasticity monitoring rather than external environment classification or phase signals.

\section{Conclusion}
\label{sec:conclusion}

This paper argued that plasticity loss is a fundamental, under-addressed bottleneck in non-stationary UAV-ECN control, and proposed PRIME to address it through internal monitoring rather than external environment modeling.
PRIME uses the shared-network batch structure of CTDE to compute team-level activation and gradient statistics, identifies neurons that are simultaneously forward-dormant and backward-silent, and reinitializes only this intersection set with an output-preserving reset.
The procedure requires no architectural changes, auxiliary losses, or phase-aware scheduling.

Direct measurement on a vanilla MAPPO controller established the three findings behind the design: dormancy accumulates and persists, forward dormancy decouples from the actual training gradient in three of four hidden layers, and per-agent dormancy decisions disagree in policy layers.
PRIME answers each in turn with periodic resets, the bidirectional intersection, and team-level aggregation.

A dynamic regret bound, obtained by extending ReSiN's single-agent tracking analysis to the team-averaged objective, shows that the perturbation energy of PRIME's resets scales with the empirical silent-subspace dimension $d_t^*$ rather than the full parameter count $d$, providing a tighter tracking guarantee than undiscriminated noise injection.
On a phase-switching UAV-ECN simulator, PRIME improves IQM return by $24.9\%$ over vanilla MAPPO and outperforms the forward-only baseline (PRIME-Forward) on return, coverage, served users, and collision rate.
Ablation studies confirm robustness to the dormancy threshold, gradient
threshold, reset period, and reset--phase alignment, and locate the
performance-critical choices in the training-gradient signal and team-level
aggregation rather than in the intervention operator; per-neuron diagnostics
validate that dormant fractions remain at $10$--$20\%$ throughout training.

Several directions remain open. The fixed reset period $F$ could be replaced by an adaptive trigger driven by online monitoring of DNF or feature rank; heterogeneous agent architectures would require a principled scheme for aggregating activation statistics across distinct sub-networks; hardware-in-the-loop validation would test robustness under real actuator latency and sensor noise; and combining silent neuron resets with complementary plasticity techniques such as layer normalization~\cite{lyle2024normalization} or spectral regularization may yield further gains.

\bibliographystyle{IEEEtran}
\bibliography{reference}

\FloatBarrier
\raggedbottom
\begin{IEEEbiographynophoto}{Wen Qiu}
received the Ph.D. degree in Co-creative Engineering from Kitami Institute of Technology, Kitami, Japan, in 2025. She is currently a Specially Appointed Researcher at Kitami Institute of Technology. Her research interests include deep reinforcement learning and emergency wireless communication networks, with a focus on intelligent resource allocation and network optimization for disaster response.
\end{IEEEbiographynophoto}
\vspace{-8pt}
\begin{IEEEbiographynophoto}{Zhiqiang He}
is currently pursuing a Ph.D. at the University of Electro-Communications, Tokyo, Japan. He received his M.S. degree in Control Science and Engineering from Northeastern University, Shenyang, China. His research interests include deep reinforcement learning and its applications. He previously worked at Baidu and InspirAI, where he developed a master-level AI for the card game Dou Di Zhu that outperformed professional players.
\end{IEEEbiographynophoto}
\vspace{-8pt}
\begin{IEEEbiographynophoto}{Wei Zhao}
(S’12-M’16) received his Ph.D. degree in the Graduate School of Information Sciences, Tohoku University. He is currently a Professor at the School of Computer Science and Technology, Anhui University of Technology. His research interests include deep reinforcement learning, edge computing, and resource allocation in wireless networks. He was the recipient of the IEEE WCSP-2014 Best Paper Award, and IEEE GLOBECOM-2014 Best Paper Award. He is a member of IEEE.
\end{IEEEbiographynophoto}
\vspace{-8pt}
\begin{IEEEbiographynophoto}{Hiroshi Masui}
received his Ph.D. in Science from Osaka University in 1998.
He currently works at the Department of Information and Communication Engineering, 
and the Vice President of Kitami Institute of Technology. 
His research interests include theoretical nuclear physics, 
public transportation analysis, cloud optimization, and data-driven science.
\end{IEEEbiographynophoto}

\end{document}


\maketitle

\renewcommand{\thefigure}{S\arabic{figure}}
\renewcommand{\thetable}{S\arabic{table}}
\renewcommand{\theequation}{S\arabic{equation}}

This document contains the appendices of the main paper. Sections are lettered A--J as referenced from the main text; figures, tables, and equations carry the prefix S.

\appendices

\section{Training Hyperparameters}
\label{app:hyperparams}

Table~\ref{tab:hyper_params} lists the complete set of training hyperparameters shared by all three methods (MAPPO, PRIME-Forward, PRIME) and the PRIME-specific settings.

\begin{table}[!t]
\renewcommand{\arraystretch}{1.3}
\caption{Training Hyperparameters}
\label{tab:hyper_params}
\centering
\begin{tabular}{l l l}
\hline
\textbf{Parameter} & \textbf{Symbol} & \textbf{Value} \\
\hline
\multicolumn{3}{l}{\emph{MAPPO (shared by all methods)}} \\
\hline
Hidden layer width & $H$ & 32 \\
Number of hidden layers & -- & 2 \\
Activation function & -- & ReLU \\
Learning rate & $\eta$ & $3 \times 10^{-4}$ \\
Weight decay & -- & $10^{-4}$ \\
Rollout length & $T_{\mathrm{roll}}$ & 2048 \\
Mini-batches per update & -- & 32 \\
PPO update epochs & $K_{\mathrm{epoch}}$ & 8 \\
Discount factor & $\gamma$ & 0.99 \\
GAE parameter & $\lambda_{\mathrm{GAE}}$ & 0.95 \\
PPO clip ratio & $\epsilon$ & 0.15 \\
Value loss coefficient & $c_v$ & 2.0 \\
Entropy coefficient & $c_e$ & $0.01$ \\
Max gradient norm & -- & 0.5 \\
Optimizer & -- & Adam ($\beta_1{=}0.9,\,\beta_2{=}0.999$) \\
Random seeds & -- & 42, 43 \\
\hline
\multicolumn{3}{l}{\emph{PRIME-specific}} \\
\hline
Dormancy threshold & $\tau_d$ & 0.5 \\
Gradient-silence threshold & $\tau_g$ & 0.08 \\
Reset period & $F$ & 200 mini-batch steps \\
Reinitialization method & -- & Kaiming uniform \\
Output-weight zeroing & -- & Yes \\
\hline
\end{tabular}
\end{table}

\section{Proof of Theorem~1}
\label{app:proof_thm1}

We first establish a lemma that quantifies the perturbation energy in the silent subspace.

\begin{lemma}[Silent Subspace Noise Energy]
\label{lem:noise}
Let $\Pi_t$ be an orthogonal projection of rank $d_t^*$.
Then $\mathbb{E}\|\Pi_t\zeta_t\|^2 = d_t^*$ for $\zeta_t \sim \mathcal{N}(0, I_d)$.
\end{lemma}

\begin{proof}
Since $\Pi_t^2 = \Pi_t = \Pi_t^\top$, we have
$\mathbb{E}\|\Pi_t\zeta_t\|^2 = \mathbb{E}[\zeta_t^\top \Pi_t \zeta_t] = \mathrm{tr}(\Pi_t) = d_t^*$.
\end{proof}

\begin{proof}[Proof of Theorem~1 of the main paper]
Define the tracking error $e_t = w_t - w_t^*$ and the drift $\Delta_t = w_{t+1}^* - w_t^*$.
Substituting the update rule~the update rule of the main paper (Section~IV-F) and using the first-order optimality condition $\nabla\mathcal{L}_t(w_t^*) = 0$:
\begin{align}
e_{t+1} &= w_{t+1} - w_{t+1}^* \notag\\
&= \underbrace{e_t - \eta\bigl(\nabla\mathcal{L}_t(w_t) - \nabla\mathcal{L}_t(w_t^*)\bigr)}_{A_t} - \Delta_t + \eta\gamma_{\mathrm{r}}\Pi_t\zeta_t.
\label{eq:error_decomp}
\end{align}
\textbf{Step 1: Contraction.}
By $\mu$-strong convexity and $L$-smoothness with $\eta \leq 1/L$, the term $A_t$ satisfies~\cite{he2025resin}:
\begin{equation}
\|A_t\|^2 \;\leq\; (1 - \mu\eta)\|e_t\|^2.
\label{eq:contraction}
\end{equation}

\textbf{Step 2: Squaring and taking expectation.}
From~\eqref{eq:error_decomp}, since $\mathbb{E}[\zeta_t] = 0$ and $\zeta_t$ is independent of $A_t$ and $\Delta_t$:
\begin{equation}
\mathbb{E}\|e_{t+1}\|^2 = \mathbb{E}\|A_t - \Delta_t\|^2 + \eta^2\gamma_{\mathrm{r}}^2 d_t^*,
\label{eq:squared}
\end{equation}
where we used Lemma~\ref{lem:noise}.
Applying the parameterized Young's inequality $\|a - b\|^2 \leq (1+\alpha)\|a\|^2 + (1+\alpha^{-1})\|b\|^2$ with $\alpha = \mu\eta/(2-\mu\eta)$, so that $1+\alpha^{-1} = 2/(\mu\eta)$ exactly:
\begin{align}
\mathbb{E}\|A_t - \Delta_t\|^2 &\leq \tfrac{2(1-\mu\eta)}{2-\mu\eta}\,\mathbb{E}\|e_t\|^2 + \tfrac{2}{\mu\eta}\|\Delta_t\|^2 \notag\\
&\leq \bigl(1 - \tfrac{\mu\eta}{2}\bigr)\,\mathbb{E}\|e_t\|^2 + \tfrac{2}{\mu\eta}\|\Delta_t\|^2,
\label{eq:young}
\end{align}
where the second inequality uses $\tfrac{2(1-\mu\eta)}{2-\mu\eta} \leq 1 - \tfrac{\mu\eta}{2}$, which holds for all $\mu\eta \in (0,1]$ and in particular for $\eta \leq 1/L \leq 1/\mu$.

\textbf{Step 3: Combining.}
Substituting~\eqref{eq:young} into~\eqref{eq:squared}:
\begin{equation}
\mathbb{E}\|e_{t+1}\|^2 \;\leq\; \bigl(1 - \tfrac{\mu\eta}{2}\bigr)\,\mathbb{E}\|e_t\|^2 + \tfrac{2}{\mu\eta}\|\Delta_t\|^2 + \eta^2\gamma_{\mathrm{r}}^2 d_t^*.
\label{eq:recursion}
\end{equation}

\textbf{Step 4: Telescoping.}
Let $\rho = 1 - \mu\eta/2$.
Unrolling~\eqref{eq:recursion} from $t = 0$ to $T-1$:
\begin{equation}
\mathbb{E}\|e_t\|^2 \;\leq\; \rho^t\,\mathbb{E}\|e_0\|^2 + \tfrac{2}{\mu\eta}\sum_{k=0}^{t-1}\rho^{t-1-k}\|\Delta_k\|^2 + \eta^2\gamma_{\mathrm{r}}^2\sum_{k=0}^{t-1}\rho^{t-1-k} d_k^*.
\end{equation}
Averaging from $t=1$ to $T$ and using $\frac{1}{T}\sum_{t=1}^{T}\rho^t \leq \frac{2}{\mu\eta T}$ and $\sum_{k=0}^{t-1}\rho^{t-1-k} \leq \frac{2}{\mu\eta}$:
\begin{equation}
\begin{aligned}
\frac{1}{T}\sum_{t=1}^{T}\mathbb{E}\|e_t\|^2 \;\leq\;
& \frac{2}{\mu\eta T}\,\mathbb{E}\|e_0\|^2 + \frac{4}{\mu^2\eta^2 T}\sum_{k=0}^{T-1}\|\Delta_k\|^2 \\
& + \frac{2\eta\gamma_{\mathrm{r}}^2}{\mu}\cdot\frac{1}{T}\sum_{t=1}^{T} d_t^*.
\end{aligned}
\end{equation}
Because the cross terms are nonnegative, $\sum_{k=0}^{T-1}\|\Delta_k\|^2 \leq \bigl(\sum_{k=0}^{T-1}\|\Delta_k\|\bigr)^2 = P_T^2$.
Collecting the first two terms yields the bound of Theorem~1.
\end{proof}

\section{Remark on Assumptions of Theorem~1}
\label{app:remark_assumptions}

Assumptions~1 and~2 (smoothness and local strong convexity of the PPO surrogate loss) are standard in the PPO tracking-error literature~\cite{schulman2017proximal, lascu2025ppo, he2025resin, qiu2026plasticity}; they hold locally within the trust region enforced by the PPO clip constraint, not globally over the full parameter space.
The additive-noise model for resets (Assumption~4) is a tractable proxy for three operations performed during each PRIME reset:
(i)~Kaiming reinitialization, whose per-neuron perturbation energy is $\mathbb{E}\|\Delta\mathbf{w}_{l,i}\|^2 = 1$ independent of layer width;
(ii)~output-weight zeroing, which makes the immediate network-output perturbation exactly zero; and
(iii)~Adam state clearing, which transiently raises the effective learning rate for reset neurons.
Because the Gaussian proxy upper-bounds (i) and ignores the perturbation-reducing effect of (ii), the bound in Theorem~1 of the main paper is conservative: it overestimates perturbation energy and thus understates the advantage of selective over blanket resets.
The operator ablation in Section~V of the main paper supports the proxy directly: replacing PRIME's reset with a matched-cadence Gaussian perturbation of the same silent set leaves Change-mode performance on par ($74.414$ versus $72.626$ IQM), consistent with the bound's dependence on the intervention subspace rather than on the specific operator.
The primary value of the theorem is qualitative: it identifies $d_t^*$ as the key quantity separating selective from blanket perturbation, a prediction the ablation experiments confirm.

\section{Normal-Mode Motivation Measurements}
\label{app:normal_motivation}

\begin{figure}[!t]
\centering
\includegraphics[width=\columnwidth]{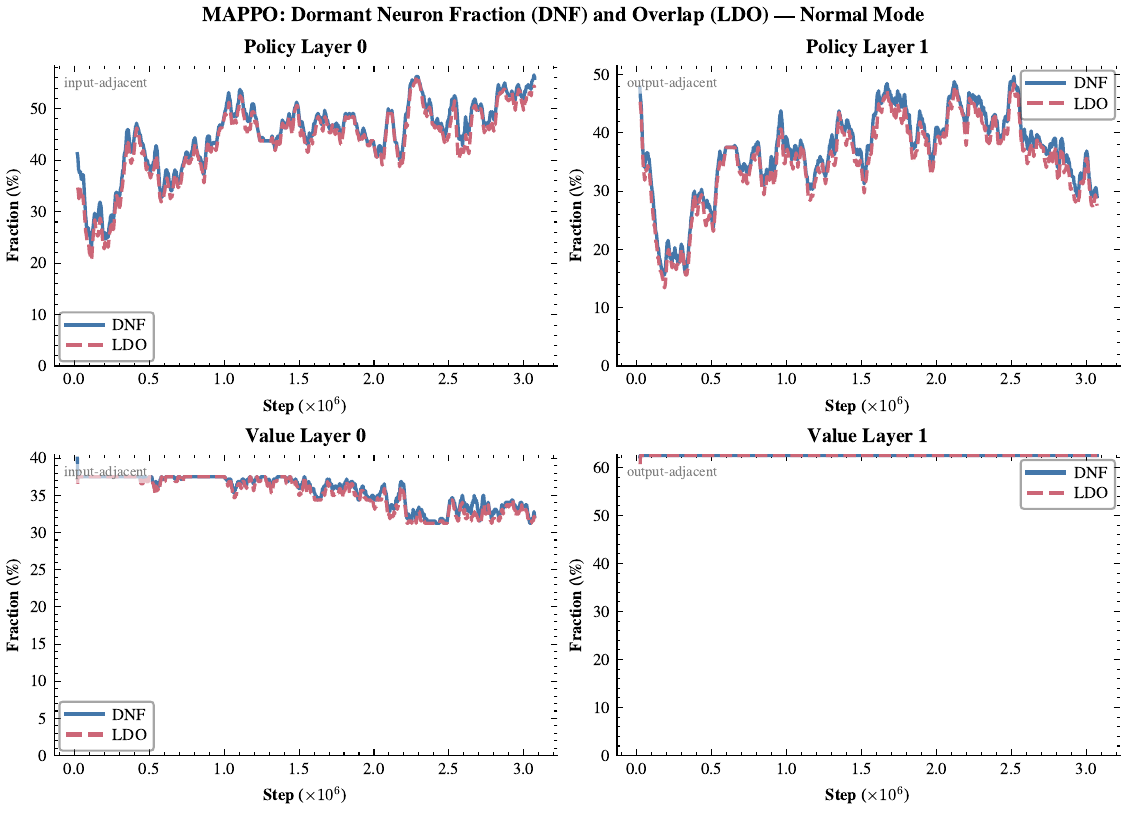}
\caption{Normal-mode DNF and LDO ($3.07{\times}10^6$ steps); cf. Fig.~1 of the main paper.}
\label{fig:motivation_normal}
\end{figure}

\begin{figure}[!t]
\centering
\includegraphics[width=\columnwidth]{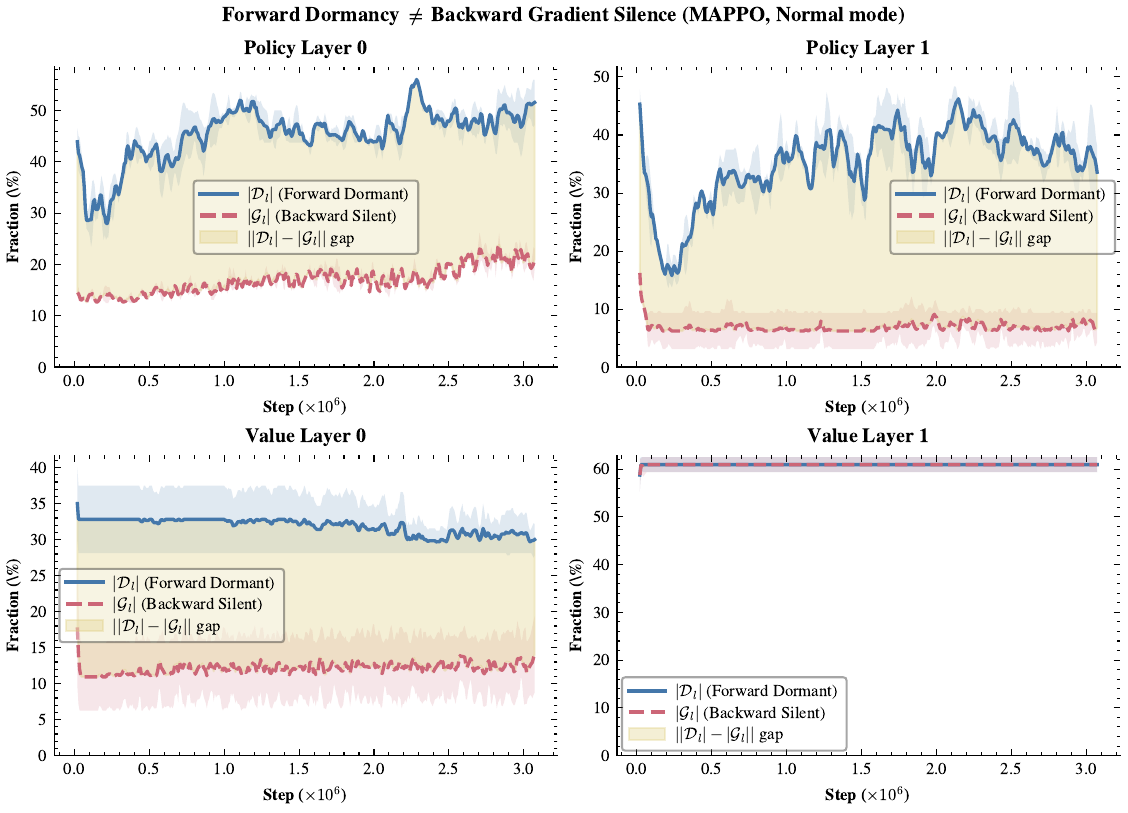}
\caption{Normal-mode forward--backward decoupling; cf. Fig.~2 of the main paper.}
\label{fig:decoupling_normal}
\end{figure}

\begin{figure}[!t]
\centering
\includegraphics[width=\columnwidth]{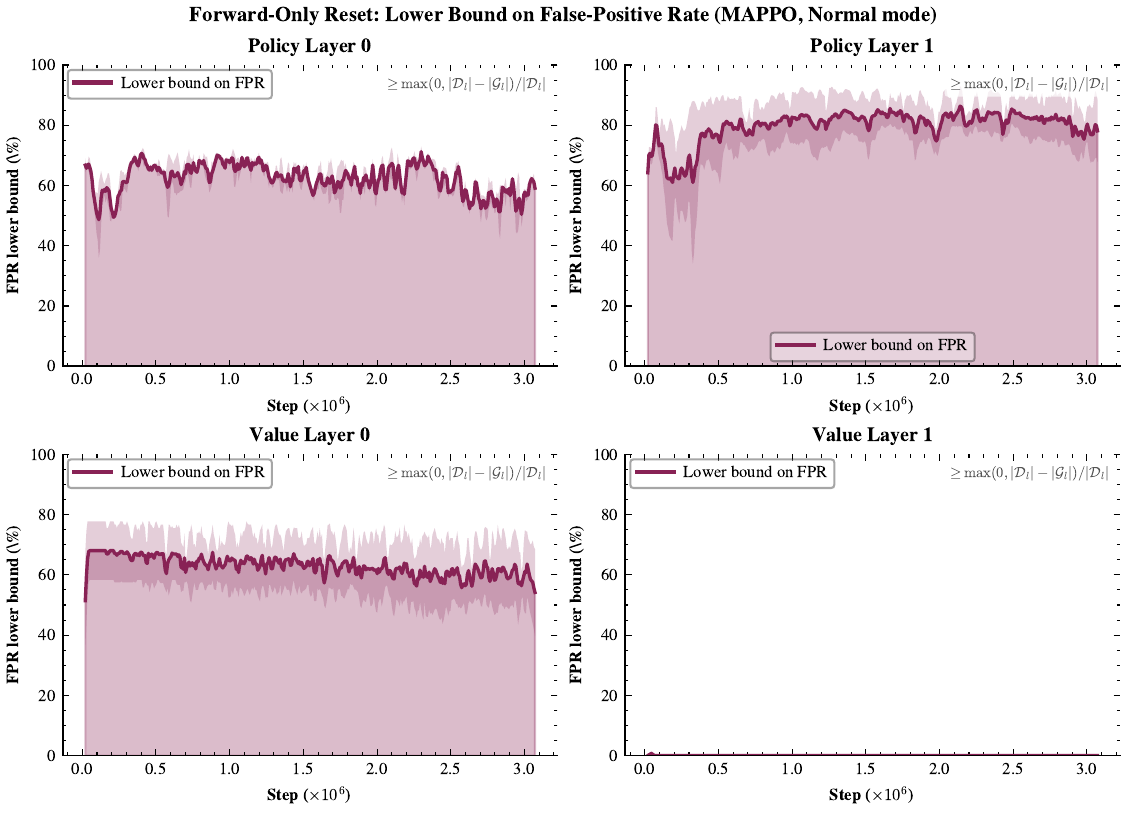}
\caption{Normal-mode false-positive lower bound; cf. Fig.~3 of the main paper.}
\label{fig:fpr_lb_normal}
\end{figure}

Figures~\ref{fig:motivation_normal}--\ref{fig:fpr_lb_normal} collect the Normal-mode counterparts of the vanilla-MAPPO measurements in Section~II of the main paper.

The qualitative picture matches Change mode: Value Layer~1 saturates near $60\%$ with DNF and LDO coinciding, the forward-dormant set exceeds the backward-silent set in the same three layers, and the false-positive lower bound occupies comparable per-layer bands, confirming that the three findings are properties of the training dynamics rather than artifacts of phase switching.

\FloatBarrier

\section{Normal-Mode Task Performance}
\label{app:normal_task}

The main text focuses on Change mode, where plasticity loss is most pronounced.
This appendix presents the corresponding Normal-mode results.

\begin{figure}[!t]
\centering
\subfloat[Training reward]{
  \includegraphics[width=0.44\columnwidth]{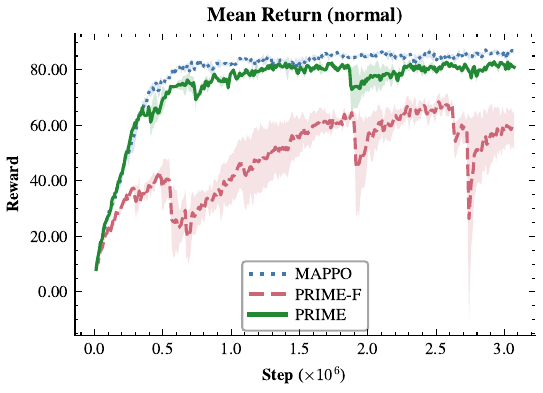}
  \label{fig:app_reward_normal}}
\hfill
\subfloat[Normalized IQM, five ECN metrics]{
  \includegraphics[width=0.47\columnwidth]{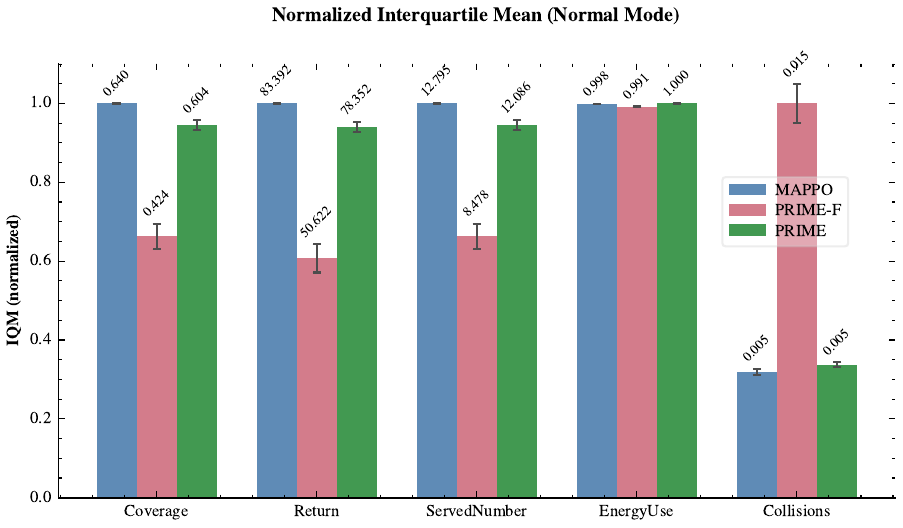}
  \label{fig:app_iqm_normal}}
\caption{Normal-mode task performance.
(a)~MAPPO attains the highest asymptotic return; PRIME converges close behind; PRIME-Forward shows instability near Step~$2.5{\times}10^6$.
(b)~MAPPO leads on return and coverage, PRIME is competitive, and PRIME-Forward trails on return.}
\label{fig:app_task_normal}
\end{figure}

Fig.~\ref{fig:app_reward_normal} shows the training reward curves.
MAPPO attains the highest asymptotic return (${\sim}\,85$), PRIME converges to ${\sim}\,80$, and PRIME-Forward shows instability near Step~$2.5{\times}10^6$ caused by an over-resetting event that destabilizes the learned trajectory policy.

Fig.~\ref{fig:app_iqm_normal} reports the normalized IQM across the five ECN metrics.
MAPPO leads on Return~($83.392$) and Coverage~($0.640$); PRIME is competitive (Return~$78.352$, Coverage~$0.604$); PRIME-Forward trails on Return ($50.622$).

\begin{figure}[!t]
\centering
\subfloat[Coverage (Normal)]{
  \includegraphics[width=0.47\columnwidth]{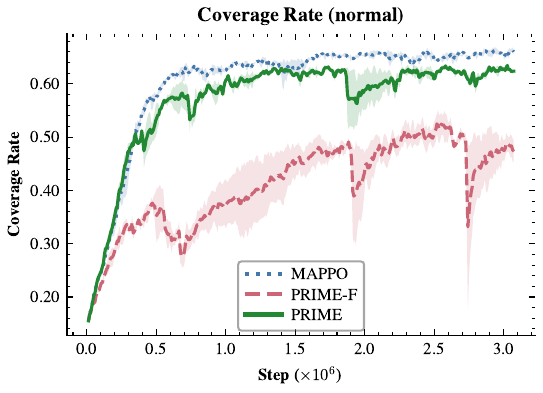}
  \label{fig:app_coverage_normal}}
\hfill
\subfloat[Served users (Normal)]{
  \includegraphics[width=0.47\columnwidth]{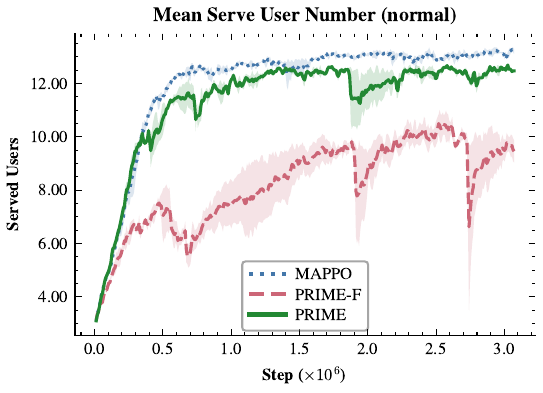}
  \label{fig:app_served_normal}}
\caption{ECN task metrics in Normal mode.
(a)~MAPPO and PRIME converge to comparable coverage (${\sim}\,0.63$--$0.65$);
PRIME-Forward lags and drops near Step~$2.5{\times}10^6$.
(b)~Served-user counts are similar for MAPPO and PRIME (${\sim}\,12$--$13$).}
\label{fig:app_ecn_normal}
\end{figure}

Fig.~\ref{fig:app_ecn_normal} shows coverage rate and served-user count.
MAPPO and PRIME converge to comparable coverage (${\sim}\,0.63$--$0.65$) and served-user counts (${\sim}\,12$--$13$), confirming that PRIME's periodic resets incur only a marginal steady-state cost.
PRIME-Forward lags behind both methods, consistent with the reward instability above.

\begin{figure}[!t]
\centering
\subfloat[Collisions (Normal)]{
  \includegraphics[width=0.47\columnwidth]{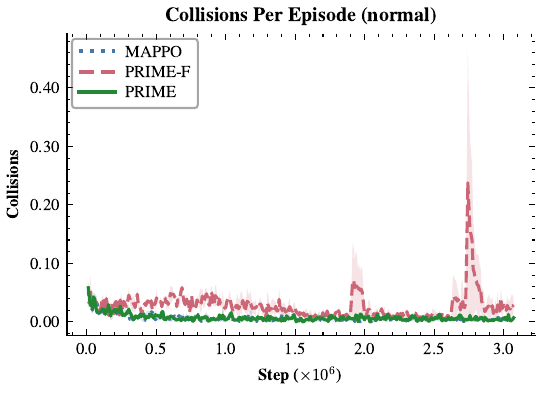}
  \label{fig:app_collisions_normal}}
\hfill
\subfloat[Energy rate (Normal)]{
  \includegraphics[width=0.47\columnwidth]{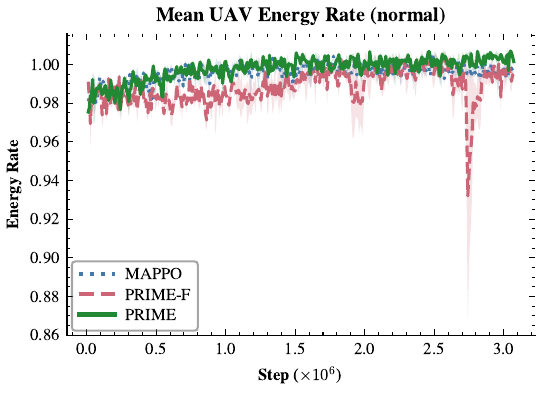}
  \label{fig:app_energy_normal}}
\caption{Safety and energy in Normal mode.
(a)~Near-zero collisions for all methods.
(b)~Energy rates are nearly identical (${\sim}\,0.99$--$1.00$).}
\label{fig:app_safety_normal}
\end{figure}

Fig.~\ref{fig:app_safety_normal} shows collision rate and energy consumption.
All three methods maintain near-zero collision rates throughout training, indicating that safety-critical trajectory constraints are satisfied regardless of the plasticity mechanism.
Energy usage rates are nearly identical across methods (${\sim}\,0.99$--$1.00$), confirming that the reinitialization procedure introduces no measurable energy overhead.

\FloatBarrier

\section{Normal-Mode Plasticity Diagnostics}
\label{app:normal_diag}

This appendix complements the Change-mode diagnostics in Section~V of the main paper with the corresponding Normal-mode indicators.

\begin{figure}[!t]
\centering
\subfloat[Dormant fraction (Normal)]{
  \includegraphics[width=0.47\columnwidth]{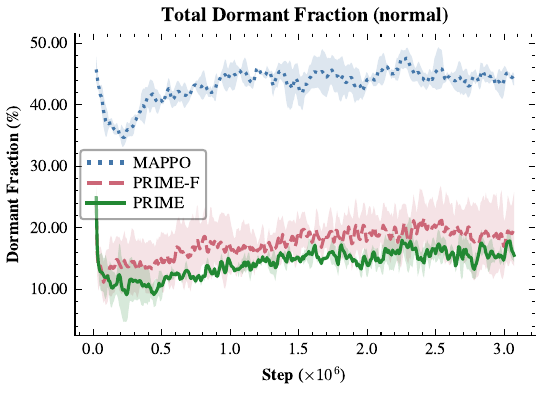}
  \label{fig:app_dormant_normal}}
\hfill
\subfloat[Feature rank (Normal)]{
  \includegraphics[width=0.47\columnwidth]{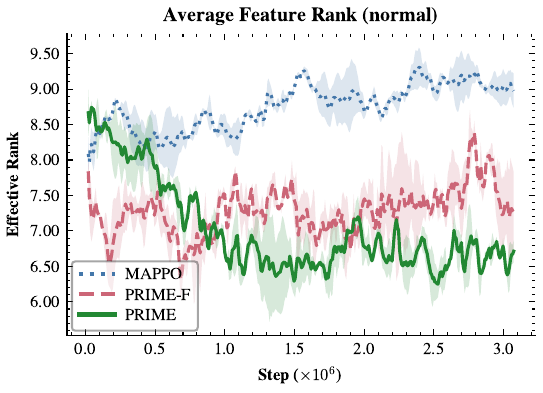}
  \label{fig:app_rank_normal}}
\caption{Plasticity diagnostics in Normal mode.
(a)~MAPPO's DNF rises to ${\sim}\,40$--$45\%$; PRIME stabilizes at ${\sim}\,10$--$18\%$.
(b)~MAPPO's higher rank (${\sim}\,9$--$10$) is inflated by noisy dormant-neuron activations.}
\label{fig:app_diag_normal}
\end{figure}

Fig.~\ref{fig:app_diag_normal}(a) shows the total dormant neuron fraction.
MAPPO's DNF rises monotonically and plateaus at ${\sim}\,40$--$45\%$, confirming that plasticity loss occurs even without phase switching.
PRIME stabilizes at ${\sim}\,10$--$18\%$.
PRIME-Forward's DNF tracks PRIME closely (${\sim}\,10$--$18\%$), confirming that periodic resets effectively suppress activation-dormancy under the stationary objective regardless of whether the gradient-silence filter is applied.
Note, however, that PRIME-Forward's near-identical dormancy curve in this stationary setting does \emph{not} translate to comparable task performance: the abrupt return drop near Step~$2.5{\times}10^6$ visible in Fig.~\ref{fig:app_ecn_normal} occurs without a corresponding dormancy spike, indicating that the resets disrupted useful representations even when the activation statistics suggested the network was healthy.

Fig.~\ref{fig:app_diag_normal}(b) shows the average feature rank.
MAPPO maintains a higher effective rank (${\sim}\,9$--$10$) than PRIME (${\sim}\,7$) throughout training.
As discussed in Section~V of the main paper, this higher rank is consistent with noisy activations from dormant neurons inflating the rank estimate without a corresponding gain in policy quality.

\begin{figure}[!t]
\centering
\includegraphics[width=0.8\columnwidth]{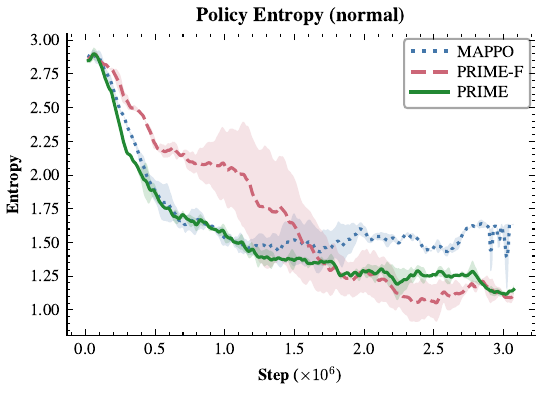}
\caption{Policy entropy in Normal mode.
All methods show decreasing entropy; the three curves remain within a narrow band of each other, with PRIME and MAPPO largely overlapping.}
\label{fig:app_entropy_normal}
\end{figure}

Fig.~\ref{fig:app_entropy_normal} shows policy entropy.
All three methods exhibit monotonically decreasing entropy as their policies converge.
PRIME's entropy curve closely tracks MAPPO's, with PRIME occasionally exceeding MAPPO in the mid-training window; the three curves remain within a narrow band of each other.

\FloatBarrier

\section{Per-Layer Dormancy Breakdown}
\label{app:per_layer}

\begin{figure}[!t]
\centering
\includegraphics[width=\columnwidth]{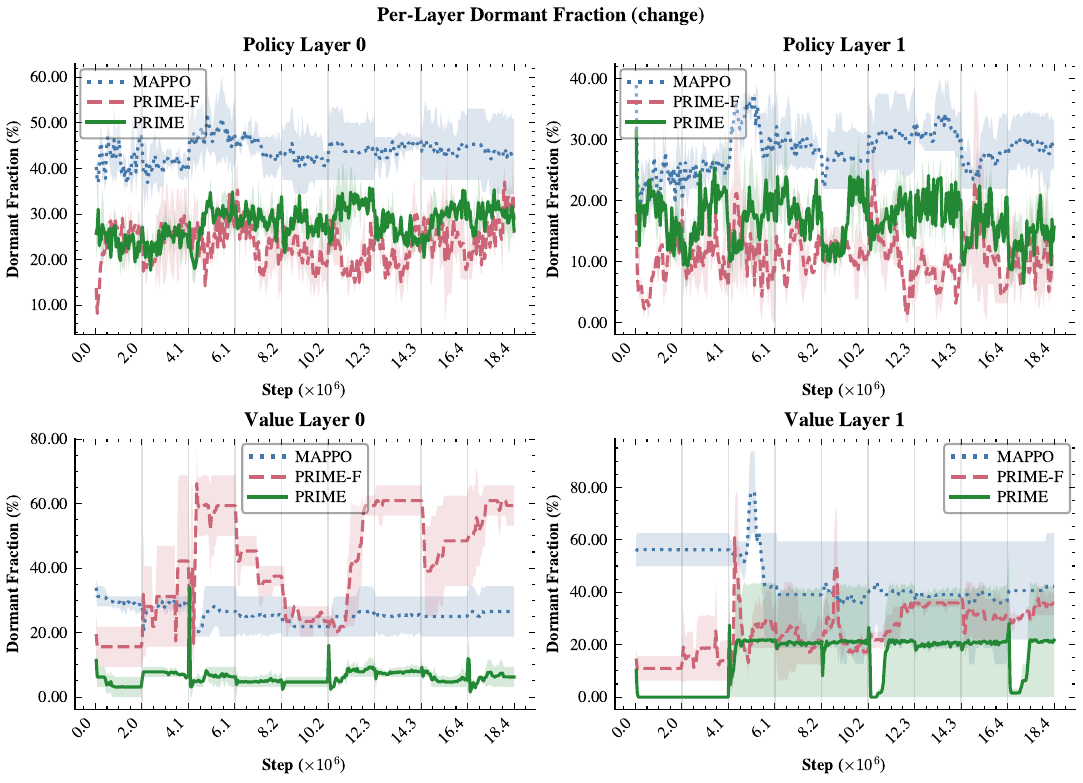}
\caption{Per-layer dormant fraction in Change mode.
Policy Layer~0 shows the highest dormancy for MAPPO (${\sim}\,25$--$55\%$).
PRIME keeps Value Layer~0 dormancy below $10\%$ and Value Layer~1 near $20\%$, reflecting the strong TD-error gradient flow.
PRIME-Forward exhibits erratic dormancy spikes in value layers from indiscriminate resets.}
\label{fig:app_per_layer}
\end{figure}

The main text reports the total dormant neuron fraction aggregated across all layers.
This appendix provides the per-layer breakdown in Change mode (Fig.~\ref{fig:app_per_layer}), which reveals that dormancy is distributed unevenly across the network.

Policy Layer~0, which directly processes the raw observation vector, exhibits the highest dormancy for MAPPO (${\sim}\,25$--$55\%$) and the largest absolute reduction under PRIME (${\sim}\,20$--$35\%$).
Early layers are most exposed to input distribution shifts at phase boundaries, which explains this pattern.
Policy Layer~1 shows lower absolute dormancy across all methods, with MAPPO remaining the highest.

The value network layers behave differently.
MAPPO's Value Layer~1 accumulates ${\sim}\,50$--$65\%$ dormant neurons, while PRIME keeps Value Layer~0 below $10\%$ and holds Value Layer~1 near $20\%$.
This asymmetry arises because the critic receives stronger, more consistent gradient signals from TD-error backpropagation, allowing PRIME's resets to integrate rapidly.
PRIME-Forward, by contrast, shows erratic dormancy spikes in the value layers (e.g., Value Layer~0 jumps to ${\sim}\,55\%$ around Step~$5{\times}10^6$).
These spikes reflect the destabilizing effect of indiscriminate forward-only resets on the critic, a failure mode that PRIME avoids through the gradient-silence filter.

\begin{figure}[!t]
\centering
\includegraphics[width=\columnwidth]{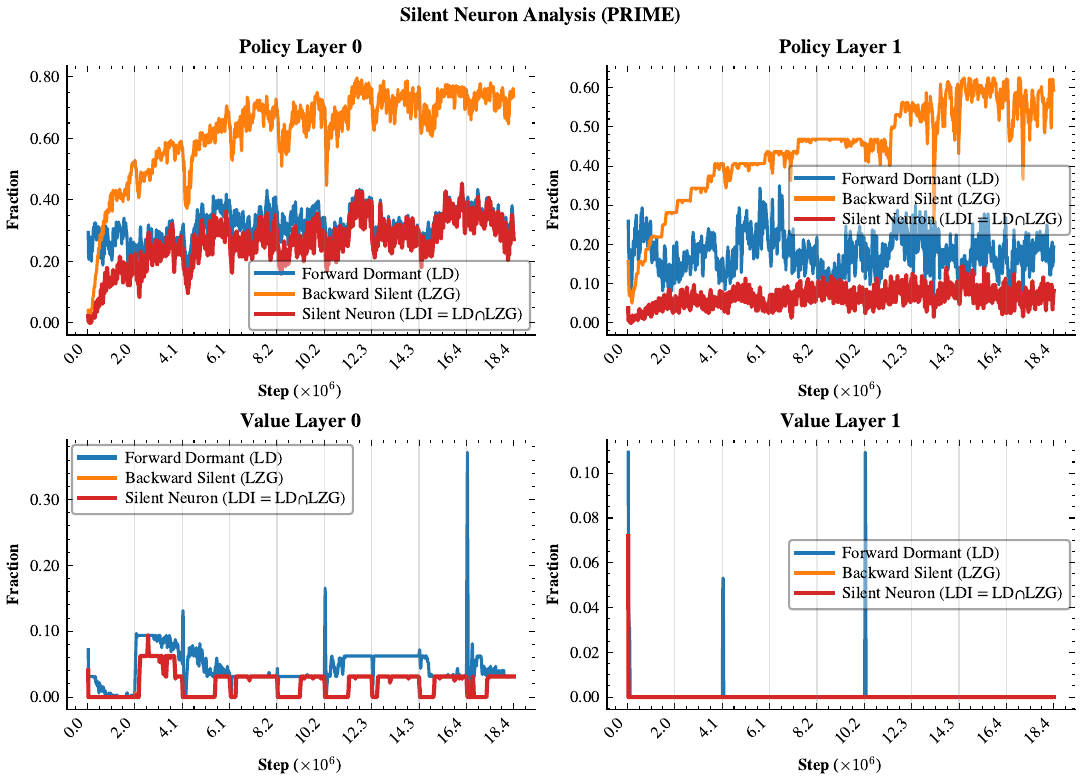}
\caption{Silent neuron decomposition for PRIME (Change mode).
Each panel shows one network layer: forward-dormant fraction ($L_D$, blue), backward gradient-silent fraction ($L_{ZG}$, orange), and their intersection ($L_{DI} = L_D \cap L_{ZG}$, red); the legend uses the plain-text labels LD, LZG, and LDI.
Policy Layer~0 exhibits $20$--$30\%$ silent neurons and Policy Layer~1 a smaller fraction; value layers show minimal dormancy.}
\label{fig:silent_neuron}
\end{figure}

Fig.~\ref{fig:silent_neuron}, referenced from Section~V of the main paper, complements this view with PRIME's per-layer decomposition into the forward-dormant, gradient-silent, and intersection sets.

\section{Energy and UAV Utilization in Change Mode}
\label{app:energy_uav}

\begin{figure}[!t]
\centering
\subfloat[Energy rate (Change)]{
  \includegraphics[width=0.47\columnwidth]{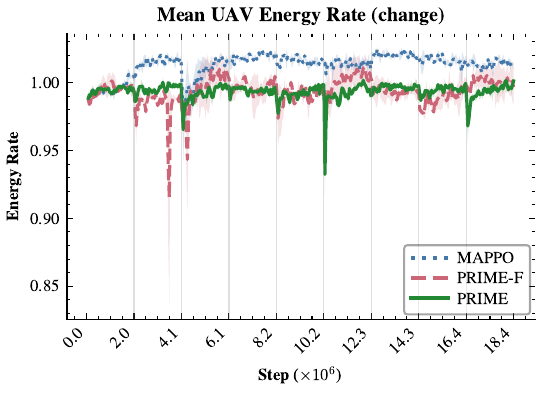}
  \label{fig:app_energy_change}}
\hfill
\subfloat[UAV using rate (Change)]{
  \includegraphics[width=0.47\columnwidth]{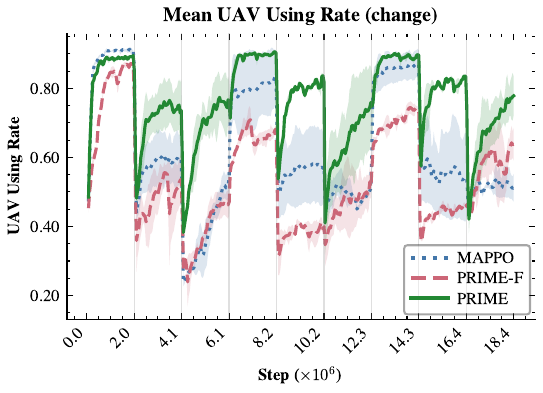}
  \label{fig:app_uav_change}}
\caption{Energy and utilization in Change mode.
(a)~Energy rates are comparable (${\sim}\,0.96$--$1.01$ apart from brief boundary dips); performance differences are not attributable to energy trade-offs.
(b)~UAV utilization rate tracks coverage: PRIME recovers fastest to ${\sim}\,0.90$.}
\label{fig:app_energy_uav}
\end{figure}

Fig.~\ref{fig:app_energy_uav} presents two supplementary Change-mode metrics that are referenced but not plotted in the main text.

Fig.~\ref{fig:app_energy_uav}(a) shows the mean UAV energy usage rate, which stays within $0.96$--$1.01$ for all three methods apart from brief transient dips at phase boundaries.
MAPPO dips furthest, briefly reaching ${\sim}\,0.86$ around Step~$10.2{\times}10^6$, but the differences are small relative to the inter-method reward and coverage gaps.
PRIME's performance advantages are therefore not attributable to differential energy expenditure.

Fig.~\ref{fig:app_energy_uav}(b) shows the UAV utilization rate, defined as the fraction of time slots during which a UAV is actively serving at least one user.
This metric qualitatively tracks the coverage rate reported in Fig.~7(a) of the main paper: PRIME achieves the fastest post-switch recovery to ${\sim}\,0.90$, while MAPPO's recovery speed degrades across later phases.
The correlation between utilization and coverage confirms that PRIME's coverage advantage stems from more effective trajectory repositioning rather than from differences in power allocation or channel conditions.

\section{Per-Layer Gradient Analysis}
\label{app:gradient}

\begin{figure}[!t]
\centering
\includegraphics[width=\columnwidth]{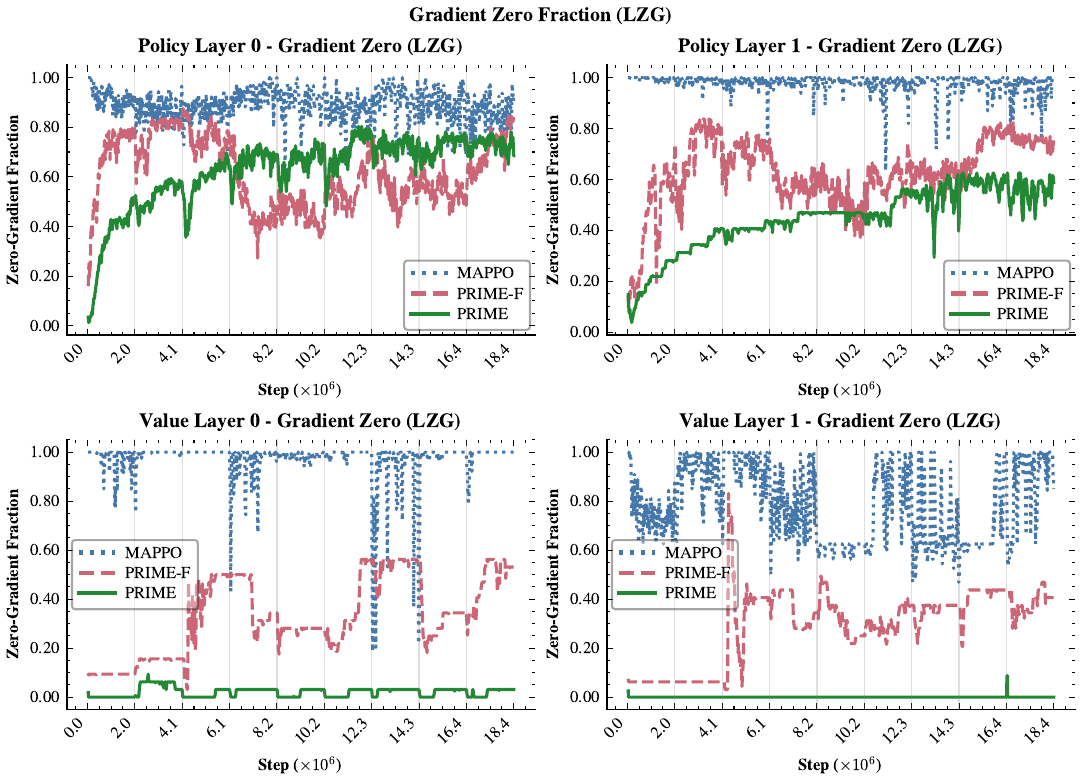}
\caption{Per-layer zero-gradient fraction (LZG) in Change mode.
Policy layers: MAPPO's LZG fluctuates ($0.6$--$1.0$); PRIME stabilizes at ${\sim}\,0.5$--$0.8$ (Layer~0) and ${\sim}\,0.3$--$0.55$ (Layer~1).
Value layers: PRIME maintains near-zero LZG ($<0.05$), confirming strong gradient flow;
MAPPO approaches $1.0$, consistent with high value-network dormancy.}
\label{fig:app_gradient_zero}
\end{figure}

The main text employs the backward-silence criterion $\hat{g}_{l,i} \leq \tau_g$ for reset decisions (Section~V of the main paper) but does not visualize the per-layer zero-gradient fraction directly.
Fig.~\ref{fig:app_gradient_zero} provides the full per-layer LZG (the layer-wise zero-gradient fraction, denoted $L_{ZG}$ in Fig.~\ref{fig:silent_neuron}) visualization in Change mode.

In the policy layers (top row), MAPPO's LZG fluctuates between $0.6$ and $1.0$, reflecting transient gradient bursts that do not translate into sustained learning.
PRIME-Forward's LZG increases over training as reset neurons begin with near-zero weights that produce small gradients.
PRIME's LZG stabilizes at ${\sim}\,0.5$--$0.8$ in Policy Layer~0 and ${\sim}\,0.3$--$0.55$ in Policy Layer~1, indicating that a meaningful fraction of policy neurons maintain active gradient flow while the silent subset is periodically refreshed.

\begin{figure}[!t]
\centering
\includegraphics[width=\columnwidth]{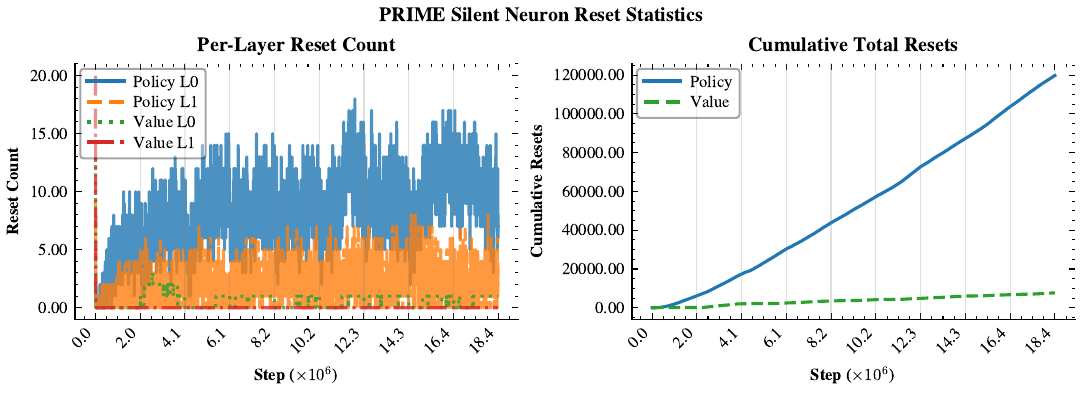}
\caption{PRIME reset statistics (Change mode).
Left: per-layer reset counts (Policy Layer~0 receives the most resets).
Right: cumulative totals show that the policy backbone accumulates ${\sim}\,1.2 \!\times\! 10^5$ resets versus ${\sim}\,5 \!\times\! 10^3$ for the critic.}
\label{fig:reset_stats}
\end{figure}

In the value layers (bottom row), the difference is pronounced.
PRIME maintains near-zero LZG ($<0.05$), confirming that virtually all value-network neurons receive strong gradient signals from TD-error backpropagation.
MAPPO's value LZG approaches $1.0$ in later phases, consistent with the high value-network dormancy documented in Appendix~\ref{app:per_layer}.
This pattern explains the asymmetric reset counts in Fig.~\ref{fig:reset_stats}: the policy network requires far more resets than the critic because the critic's gradient flow naturally resists dormancy, with Policy Layer~0 receiving ${\sim}\,5$--$20$ resets per period.

\section{Multi-Metric Reset Period Ablation}
\label{app:ablation_multi}

The main text reports reset period sensitivity using IQM return as the sole metric (Fig.~10(a) of the main paper(a)).
This appendix extends the ablation to coverage rate and dormant neuron fraction to verify that PRIME's robustness to $F$ holds across the evaluation suite.

\begin{figure}[!t]
\centering
\includegraphics[width=0.95\columnwidth]{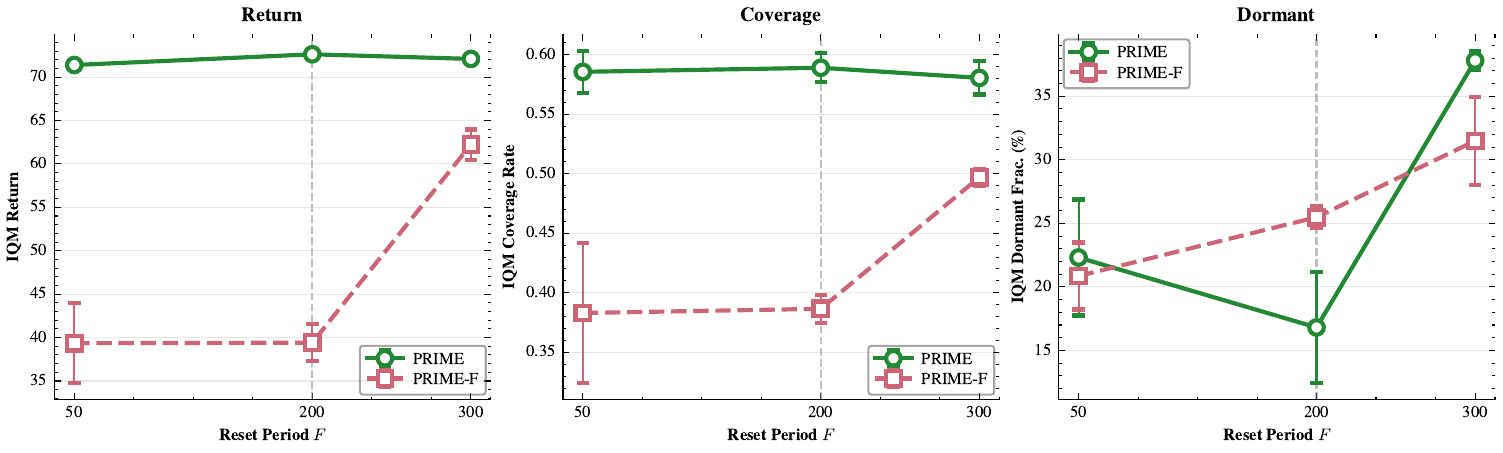}
\caption{Multi-metric reset period ablation (Change mode, $\tau_d{=}0.5$, $\tau_g{=}0.08$).
Left: IQM return; Center: coverage rate; Right: dormant fraction.
PRIME is stable across all $F$ values and metrics;
PRIME-Forward varies with reset frequency.}
\label{fig:app_ablation_multi}
\end{figure}

As shown in Fig.~\ref{fig:app_ablation_multi}, PRIME's IQM return and coverage are stable across $F \in \{50, 200, 300\}$, varying by less than $2\%$ on return.
The dormant fraction is more sensitive: it stays near $17$--$22\%$ for $F\in\{50,200\}$ but rises to ${\sim}\,38\%$ at $F{=}300$, where the inter-reset interval is long enough for dormant neurons to accumulate before the next detection sweep.
This elevated dormancy at $F{=}300$ nevertheless produces no measurable performance loss, confirming that the bidirectional silent set is selective enough that even temporarily inflated dormancy between resets remains harmless.
PRIME-Forward, in contrast, is sensitive on all three metrics: its return and coverage improve with larger $F$ (less frequent over-resetting causes less disruption), while its dormant fraction increases monotonically with $F$ (more frequent resets mechanically reduce dormancy but destroy useful representations in the process).

These results reinforce the conclusion from Section~V of the main paper: the bidirectional intersection criterion, not the specific choice of $F$, is the primary factor behind PRIME's task-level robustness.

\bibliographystyle{IEEEtran}
\bibliography{reference}